\documentclass[12pt]{article}
\usepackage[dvipsnames]{xcolor}

\usepackage{fix-cm}
\usepackage{amsmath}
\usepackage{graphicx}
\usepackage{enumerate}
\usepackage[numbers]{natbib}
\usepackage[nottoc]{tocbibind}

\usepackage{url} 

\newcommand{\blind}{1}

\usepackage[margin=0.99in]{geometry}

\newcommand{\blue}{\textcolor{black}}
\usepackage{graphicx}
\usepackage{centernot}
\usepackage{lipsum}
\usepackage{etoolbox}
\usepackage{setspace}
\usepackage{dlfltxbcodetips,bbm,mathtools,amsfonts,amsmath,amssymb,fancybox,graphicx,bm,latexsym}
\usepackage{titletoc}
\usepackage{subfigure}
\usepackage[subfigure]{tocloft}

\usepackage[mathscr]{eucal}
\newcommand{\hide}[1]{}

\newcommand{\mnorm}[1]{\left|\!\left|\!\left|#1\right|\!\right|\!\right|}

\newcommand\orth{\protect\mathpalette{\protect\independenT}{\perp}}
\def\independenT#1#2{\mathrel{\rlap{$#1#2$}\mkern2mu{#1#2}}}
\usepackage{url} 
\usepackage{hyperref}

\usepackage{multirow}

\allowdisplaybreaks

\usepackage{array}
\newcolumntype{H}{>{\setbox0=\hbox\bgroup}c<{\egroup}@{}}

\usepackage[mathscr]{eucal}
\def\independenT#1#2{\mathrel{\rlap{$#1#2$}\mkern2mu{#1#2}}}
\usepackage{dlfltxbcodetips,bbm,mathtools,subfigure,amsfonts,amsmath,amssymb,fancybox,graphicx,bm,latexsym}

\usepackage{fancyhdr}
\usepackage{tikz}
\usetikzlibrary{bayesnet}
\usetikzlibrary{fit,positioning}
\newcommand{\obs}{latent}
\newcommand{\lat}{obs}

\usepackage[tikz]{bclogo}

\newcommand{\bt}{\begin{bclogo}[couleur={rgb:orange,0;yellow,0;white,1},arrondi=0.1,logo=\bcplume,ombre=true]}
\newcommand{\et}{\end{bclogo}\s}
\newcommand{\btt}{\begin{box}}
\newcommand{\ett}{\end{box}}

\newcommand{\btheorem}{\begin{bclogo}[couleur={rgb:orange,0;yellow,0;white,1},arrondi=0.1,logo=\bcplume,ombre=true]{Theorem}}
\newcommand{\ettheorem}{\end{bclogo}}

\newcommand{\bsh}{\begin{bclogo}[couleur={rgb:orange,0;yellow,0;white,1},arrondi=0.1,logo=\bcpanchant,ombre=true]}
\newcommand{\esh}{\end{bclogo}}

\DeclareMathOperator*{\argmax}{arg\,max}

\newcommand{\benum}{\begin{enumerate}}
\newcommand{\eenum}{\end{enumerate}}

\newcommand{\bq}{\begin{quote}\em}
\newcommand{\eq}{\end{quote}}
\newcommand{\bbq}{\begin{quote}\bf\em}
\newcommand{\ebq}{\end{quote}}

\newcommand{\ind}{\msim\limits^{\mbox{\tiny ind}}}
\newcommand{\iid}{\msim\limits^{\mbox{\tiny iid}}}

\newcommand{\mR}{\mathbb{R}}

\newcommand{\mbR}{\mathbb{R}}

\newcommand{\mbI}{\mathbb{I}}

\newcommand{\mbE}{\mathbb{E}}

\newcommand{\mbP}{\mathbb{P}}

\newcommand{\mcN}{\mathcal{N}}

\newcommand{\ba}{\begin{array}{llllllllll}}
\newcommand{\ea}{\end{array}}
\newcommand{\bea}{\begin{equation}\begin{array}{llllllllll}}
\newcommand{\eea}{\end{array}\end{equation}}

\newcommand{\beno}{\begin{equation}\begin{array}{llllllllll}\nonumber}
\newcommand{\be}{\begin{equation}\begin{array}{llllllllll}}
\newcommand{\ee}{\end{array}\end{equation}}
\newcommand{\bi}{\begin{itemize}}
\newcommand{\ei}{\end{itemize}}
\newcommand{\ben}{\begin{enumerate}}
\newcommand{\een}{\end{enumerate}}

\newcommand{\dsum}{\displaystyle\sum\limits}

\newcommand{\mV}{\mathscr{V}}

\newcommand{\s}{\vspace{0.25cm}}
\newcommand{\bx}{\bm{x}}

\newcommand{\bX}{\bm{X}}
\newcommand{\mX}{\mathscr{X}}

\newcommand{\mC}{\mathscr{C}}
\newcommand{\mY}{\mathscr{Y}}

\newcommand{\mZ}{\mathscr{Z}}

\newcommand{\bY}{\bm{Y}}
\newcommand{\bC}{\bm{C}}
\newcommand{\bc}{\bm{c}}

\newcommand{\by}{\bm{y}}

\newcommand{\bu}{\bm{u}}
\newcommand{\bD}{\bm{D}}
\newcommand{\mU}{\mathscr{U}}
\newcommand{\bU}{\bm{U}}

\newcommand{\bz}{\bm{z}}
\newcommand{\bZ}{\bm{Z}}

\newcommand{\msim}{\mathop{\rm \sim}}

\newcommand{\bI}{\bm{I}}

\newcounter{comment}

\newcounter{example}

\newcounter{counterexample}

\newcounter{condition}
\newenvironment{condition}[1][]{\refstepcounter{condition}\par\medskip\noindent%
\textbf{Condition~\thecondition #1}. \rmfamily}{\medskip}

\newcounter{theorem}
\newenvironment{theorem}[1][]{\refstepcounter{theorem}\par\medskip\noindent%
\textbf{Theorem~\thetheorem #1}. \rmfamily}{\medskip}

\newcounter{proposition}

\newcounter{result}

\newcounter{proof}

\newcounter{tproof}
\newcounter{corollary}

\newcounter{cproof}
\newcounter{lemma}

\newcounter{com}

\newcounter{lproof}
\newcounter{assumption}

\newif\ifmydraft
\mydraftfalse

\ifmydraft
  
  \pagecolor{black!20}
\else
  
\fi

\ifmydraft
  
  \pagecolor{black!20}
\fi

\usepackage{multicol}

\makeatletter
\newcommand*{\deq}{\mathrel{\rlap{%
  \raisebox{0.3ex}{$\m@th\cdot$}}%
  \raisebox{-0.3ex}{$\m@th\cdot$}}=}
\makeatother

\usepackage{array}

\newcommand{\blues}{\color{black}}
\date{}

\newcommand{\mcF}{\mathscr{F}}

\usepackage{bbm}
\usetikzlibrary{fit,positioning,shapes,backgrounds,calc}
        
\tikzset{node/.style={circle, fill=white, draw, minimum size=.85cm, text = black, inner sep=1pt, thick}}

\usepackage{tikz}
\tikzstyle{const} = [rectangle, inner sep=0pt, node distance=1]
\tikzstyle{inf} = [latent,fill=red]

\usetikzlibrary{arrows}
\usetikzlibrary{positioning}

\newcommand{\pa}{\mbox{pa}}

\usetikzlibrary{fit,positioning,shapes,backgrounds,calc}

\tikzset{node/.style={circle, fill=white, draw, minimum size=.85cm, text = black, inner sep=1pt, thick}}

\usepackage{tikz}
\tikzstyle{const} = [rectangle, inner sep=0pt, node distance=1]
\tikzstyle{inf} = [latent,fill=red]

\usetikzlibrary{arrows}
\usetikzlibrary{positioning}

\numberwithin{equation}{section}
\numberwithin{theorem}{section}

\begin{document}

\def\spacingset#1{\renewcommand{\baselinestretch}%
{#1}\small\normalsize} \spacingset{1}


\if1\blind
{
  \title{\bf Causal Inference Under Network Interference}
  \author{Subhankar Bhadra\\
    Michael Schweinberger
    \\
    Department of Statistics, The Pennsylvania State University}
  \maketitle
} \fi

\if0\blind
{
  \bigskip
  \bigskip
  \bigskip
  \begin{center}
  {\LARGE\bf Causal Inference Under Network Interference}
\end{center}
  \medskip
} \fi

\bigskip

\begin{abstract}
We review the growing body of research on the challenging topic of interference in causal inference. Interference arises when the treatment assignments of connected units affect the outcomes of other units directly (via spillover) or indirectly (via contagion). Examples include public health, economic, political, or educational interventions, as well as advertising on social media. Adapting the Neyman-Rubin potential outcome framework, we review selected causal effects under interference and discuss identification of causal effects based on experimental and observational data. We then review statistical tests for interference and causal effects, design- and model-based estimators for causal effects, and methodological challenges, including variance reduction, unknown interference graphs, and misspecified exposure mappings. We conclude with a discussion of the scope of causal conclusions based on fixed and random potential outcomes and interference graphs. While the fixed potential outcome framework is sufficient in many applications, the random potential outcome framework can be useful when the data fall short. Examples include causal inference for heterogeneous treatment effects and external validity, including generalizability and transportability. 
\end{abstract}

\noindent%
\blue{{\it Keywords:} 
Experimental and observational data,
fixed and random potential outcomes,
causal effects,
scope of causal conclusions}

\pagestyle{plain}
\spacingset{1.3} 

\newpage

\tableofcontents

\section{Introduction}
\label{sec:introduction}

Causal inference has witnessed a surge of interest.
The pioneering work on causal inference by Neyman \citep{NeDaSp90}, 
Rubin \citep{Ru74}, 
Pearl \citep{Pe09}, 
and others assumes that there is no interference,
i.e.,
the outcomes of units are not affected by the treatment assignments of other units \citep{cox1958planning}.
In the interconnected and interdependent world of the twenty-first century, 
the assumption that the outcomes of units are unaffected by others may be violated,
and interventions can have both intended and unintended consequences \cite[e.g.,][]{sobel2006randomized,hong2006evaluating,cai2015social,Fr17,viviano2024policy,halloran2016dependent,LeOg21,lee2023finding}.

A simple example,
which we will use as an informal example for illustrating interference,
is advertising on social media:
advertisers may be interested in the effect of targeting teenagers on social media with advertisements of designer clothes on purchases of designer clothes.
Interference is of interest to advertisers,
because the effect of advertisements can spill over from targeted to non-targeted teenagers and can hence lift purchases of designer clothes among targeted and non-targeted teenagers.
Other examples of interference include
\bi 
\item the effect of interventions on the spread of contagious diseases and public health \citep[e.g.,][]{halloran2016dependent,lee2023estimating,lee2023finding};
\item the effect of economic and financial interventions on economic welfare \citep[e.g.,][]{cai2015social,Fr17,viviano2024policy};
\item the analysis of spatio-temporal phenomena \citep{reich2021review}, 
including the effect of conflict on forest loss \citep{ChBaKuMaPe22} and the effect of airstrikes on insurgencies \citep{papadogeorgou2022causal};
\item the effects of educational and socio-economic interventions \citep[e.g.,][]{hong2006evaluating,sobel2006randomized,ClHa24}.
\ei

The study of causal inference under interference started in its earnest with \citet{HuHa08} and other pioneering works \citep[e.g.,][]{halloran1995causal,sobel2006randomized,rosenbaum2007interference,tchetgen2012causal,aronow2012general,manski2013identification,ugander2013graph,OgVW14,liu2014large,liu2016inverse,OgVw17,aronow2017estimating}.
We provide a comprehensive review of the growing body of research on the challenging topic of interference in causal inference.
Other,
less recent reviews can be found in \citet[focusing on public health]{halloran2016dependent} and \citet[focusing on causal diagrams]{OgVW14}.
Two recent,
more specialized treatments are concerned with spatial interference \cite{reich2021review} and bipartite interference \cite{zigler2020bipartite}.

\subsection{Related Literature Not Covered}
\label{sec:literature.not.covered}

The two main approaches to causal inference without interference are based on Pearl's causal framework and Neyman's and Rubin's causal framework \citep[e.g.,][]{pearl2012causal,RiRo13,ImRu15,PeJaSc17,DiLi18,hernan2020causal,Larsen2024ab}:
\bi 
\item Pearl's causal framework is based on potential interventions in a directed acyclic graph representing causal relationships among variables \citep[e.g.,][]{Pe09}.
\vspace{.05cm}
\item Neyman's and Rubin's causal framework is based on potential outcomes \citep[e.g.,][]{NeDaSp90,Ru74,ImRu15,DiLi18}.
\ei
We review advances in causal inference under interference using the Neyman-Rubin framework.
We focus on interference in the general setting and do not discuss works concerned with special forms of interference, 
such as partial interference restricted to non-overlapping subpopulations \citep[e.g.,][]{HuHa08,tchetgen2012causal,liu2014large,rigdon2015exact,liu2016inverse,kang2016peer,basse2018analyzing,baird2018optimal,shpitser2017modeling,li2019prediction, papadogeorgou2019causal, imai2021causal},
interference arising from marketplace competition 
\citep[e.g.,][]{blake2014marketplace,liu2021trustworthy,bojinov2023design,hu2025switchback,Larsen2024ab,munro2025market},
spatial interference \cite{reich2021review},
and bipartite interference \cite{zigler2020bipartite}.
We also do not discuss the related but distinct literature on peer influence \citep[e.g.,][]{manski2013identification,bramoulle2009identification,OgVW14,
ShTh11,
KoDa22,
MFSh23,
WuLe23,
hayes2024peer,
nath2025identifying} and network-induced dependence \citep[e.g.,][]{li2019prediction,le2022,lei2024}.

\subsection{Disclaimers}
\label{sec:disclaimer}

We would like to make two additional disclaimers.

First,
the survey covers the entire field of causal inference under interference,
without making assumptions about the data-generating process.
That said,
we sometimes use examples to introduce ideas in simple settings (e.g., linear models capturing interference).
We do not endorse linear models or other models,
but we believe that examples are instructive and benefit novices,
which helps grow the community of researchers.

Second,
there are two lines of research on interference,
one of them based on fixed potential outcomes and the other based on random potential outcomes.
Important advances have been made in both lines of research.
We therefore cover both of them and attempt to build bridges between them,
including in Sections \ref{sec:from.fixed.to.random},
\ref{sec:capturing.contagion},
and \ref{fixed-to-random}.
Having said that,
there is more work on fixed potential outcomes than random potential outcomes.

\section{Causal Framework}
\label{sec:causal.framework}

We review advances in causal inference under interference through the lens of the Neyman-Rubin framework.
We first introduce the Neyman-Rubin framework in Section \ref{sec:neyman.rubin.with.interference}.
We then explain how statements based on fixed potential outcomes can be related to statements based on random potential outcomes in Section \ref{sec:from.fixed.to.random}.
Throughout the remainder of the survey,
we assume that observations of treatment assignments and outcomes at a single time point are available,
although we briefly mention recent advances in causal inference based on temporal and spatio-temporal data in Sections \ref{subsec:temporal} and \ref{sec:spatial.interference},
 respectively.

\subsection{Neyman-Rubin Framework}
\label{sec:neyman.rubin.with.interference}

In the simplest case,
the classic Neyman-Rubin framework assumes that each unit $i$ has two potential outcomes:
$Y_i(0)$ is the potential outcome that would be observed if $i$ were assigned to the control group, and $Y_i(1)$ is the potential outcome that would be observed if $i$ were assigned to the treatment group in an experiment or quasi-experiment.
The Neyman-Rubin framework has at least two notable properties.
First, 
the difference in potential outcomes $Y_i(1) - Y_i(0)$ can be attributed to the treatment of unit $i$,
provided that the potential outcomes $Y_i(0)$ and $Y_i(1)$ of unit $i$ are not affected by the treatment assignments of other units.
Second,
causal inference can be framed as a missing data problem,
because $Y_i(0)$ and $Y_i(1)$ cannot be both observed,
so one of them is missing.

The classic Neyman-Rubin framework makes the Stable Unit Treatment Value Assumption (SUTVA) \citep{rubin1986comment},
which assumes that the potential outcomes $Y_i(0)$ and $Y_i(1)$ of units $i$ are not affected by the treatment assignments of other units.
In other words,
the classic Neyman-Rubin framework rules out interference.
To accommodate interference in the Neyman-Rubin framework,
we allow the potential outcomes $Y_i(\bx)$ of units $i$ to be functions of the treatment assignments $\bx \in \mX$ of all units.
We denote the potential outcomes of all units by $\bY(\bx)$ and define $\bY(\cdot)\coloneqq\{\bY(\bx):\, \bx \in \mX\}$.

\subsection{From Fixed to Random Potential Outcomes}
\label{sec:from.fixed.to.random}

In the literature on causal inference under interference,
at least two lines of research can be distinguished:
one of them views potential outcomes as fixed,
while the other considers them to be random.
To distinguish them,
we use lowercase letters $\by(\cdot)$ when potential outcomes are fixed,
and uppercase letters $\bY(\cdot)$ when potential outcomes are random.
Statements based on fixed potential outcomes $\by(\cdot)$ can be related to statements based on random potential outcomes $\bY(\cdot)$,
because statements concerned with $\by(\cdot)$ can be interpreted as conditional statements given $\bY(\cdot) = \by(\cdot)$.
Since important advances have been made in both branches of the literature,
we review advances in causal inference under interference based on fixed and random potential outcomes side by side.

\subsection{Notation}
\label{sec:notation}

We consider $N \geq 2$ units $1, \ldots, N$ divided into a control group and a treatment group in an experiment or quasi-experiment.
The treatment assignment of unit $i$ is denoted by $X_i$,
where $X_i \coloneqq 0$ indicates that $i$ is assigned to the control group and $X_i \coloneqq 1$ indicates that $i$ is assigned to the treatment group.
The $N$-vector of treatment assignments $X_1, \ldots, X_N$ is denoted by $\bX \in \mX \subseteq \{0, 1\}^N$;
note that $\mX$ may be a proper subset of $\{0, 1\}^N$.
Throughout,
we assume that the treatment assignments $\bX$ are random in both experimental and quasi-experimental settings,
in line with the broader literature on causal inference \citep[e.g.,][p.~5]{rosenbaum2020design}.
The Neyman-Rubin framework assumes the existence of potential outcomes $Y_i(\bx) \in \mR$ that would be observed if the units received treatment assignments $\bx \in \mX$.
Since there are $2^N$ possible treatment assignments $\bx \in \mX$,
each unit $i$ has $2^N$ potential outcomes $Y_i(\bx)$ ($\bx \in \mX$).
Throughout,
$Y_j(x_i; \bx_{-i})$ denotes the potential outcome of unit $j$ when the treatment assignment of unit $i$ is $x_i$ and the treatment assignments of all other units are $\bx_{-i}$,
where $i$ can be equal to $j$.
The $N$-vector of potential outcomes $Y_1(\bx), \ldots, Y_N(\bx)$ is denoted by $\bY(\bx) \in \mbR^N$ ($\bx \in \mX$),
and the set of all potential outcomes is denoted by $\bY(\cdot)\coloneqq\{\bY(\bx):\,\bx\in\mX\} \in \mY(\cdot) \subseteq \mR^{N \times 2^N}$.
The $N$-vector of observed outcomes $Y_1, \ldots, Y_N$ is denoted by $\bY \in \mbR^N$.
According to the conventional consistency assumption of causal inference under interference \citep[e.g.,][]{TTFuSh21},
the observed outcomes $\bY$ are related to the potential outcomes $\bY(\bx)$ by
\be
\label{consistency.assumption}
\bY 
= 
\dsum_{\bx \in \mX} \bY(\bx)\, \mbI(\bX = \bx),
\ee
where $\mbI(\cdot)$ is an indicator function,
which is $1$ if its argument is true and is $0$ otherwise.
The vectors $\bX$ and $\bY$ excluding $X_i$ and $Y_i$ are denoted by $\bX_{-i}$ and $\bY_{-i}$,
respectively.
If $\bY(\cdot)$ is random,
we denote probability measures and expectations with respect to $\bX$, $\bY(\cdot)$, $(\bX, \bY(\cdot))$
by
$\mbP_{\mX}$,
$\mbP_{\mY(\cdot)}$,
$\mbP_{\mX,\mY(\cdot)}$
and
$\mbE_{\mX}$,
$\mbE_{\mY(\cdot)}$,
$\mbE_{\mX,\mY(\cdot)}$,
respectively.
The $N$-vectors of $0$'s and $1$'s are denoted by $\bm{0}_N$ and $\bm{1}_N$,
respectively.
The $N \times N$-identity matrix is $\bm{I}_N$.
The $\ell_2$-norm of vectors $\bm{a} \in \mR^d$ ($d \geq 1$) is denoted by $|\!|\bm{a}|\!|_2 \coloneqq \sqrt{\sum_{i=1}^d a_i^2}$,
and the spectral norm of matrices $\bm{A} \in \mR^{d \times d}$ is denoted by $\mnorm{\bm{A}}_2 \coloneqq \sup_{\bm{u} \in \mR^d:\; |\!|\bm{u}|\!|_2=1}\, |\!|\bm{A}\, \bm{u}|\!|_2$.

\section{Interference}
\label{sec:interference}

Interference arises when the treatment assignments of units affect the potential outcomes of other units directly (via spillover) or indirectly (via contagion),
as illustrated in Figures \ref{graph.ts.os} and \ref{graph.contagion}.
We review possible forms and representations of interference in Section \ref{sec:interference}:
\bi
\item We first introduce spillover and contagion in Sections \ref{sec:treatment.spillover} and \ref{sec:outcome.spillover},
respectively.
\vspace{.05cm}
\item We then explain in Section \ref{sec:causation.versus.correlation} that,
based on data at a single time point, 
contagion cannot be identified as a causal process,
but it can be accommodated by capturing statistical dependence among outcomes.
\vspace{.05cm}
\item Section \ref{sec:capturing.contagion} demonstrates how dependence induced by contagion can be represented in the fixed and random potential outcome frameworks.
\vspace{.05cm}
\item Section \ref{sec:other.sources} discusses other sources of dependence,
including unobserved variables.
\vspace{.05cm}
\item Section \ref{sec:interference.representations} represents interference via interference graphs and exposure mappings.
\ei 
Since statements based on fixed and random potential outcomes $\by(\cdot)$ and $\bY(\cdot)$ can be related to each other (Section \ref{sec:from.fixed.to.random}),
we discuss interference based on fixed and random potential outcomes $\by(\cdot)$ and $\bY(\cdot)$ side by side.

\begin{figure}
\begin{tabular}{ccc}
      \hspace{0.5cm} \mbox{}\tikz{ %
        \node[\obs] (y1) {$Y_1$} ; %
        \node[\obs, right=of y1] (y2) {$Y_2$} ; %
        \node[\obs, below=of y1, yshift=0cm] (h1) {$X_1$} ; %
        \node[\obs, below=of y2, yshift=0cm] (h2) {$X_2$} ; %
        \edge[color=black, line width=1pt]{h1} {y1} ; %
        \edge[color=black, line width=1pt]{h2} {y2} ; %
      }
& \hspace{1.5cm} \mbox{}
      \tikz{ %
        \node[\obs] (y1) {$Y_1$} ; %
        \node[\obs, right=of y1] (y2) {$Y_2$} ; %
        \node[\obs, below=of y1, yshift=0cm] (h1) {$X_1$} ; %
        \node[\obs, below=of y2, yshift=0cm] (h2) {$X_2$} ; %
        \edge[color=black, line width=1pt]{h1} {y1} ; %
        \edge[color=MidnightBlue, line width=1pt]{h1} {y2} ; %
        \edge[color=black, line width=1pt]{h2} {y2} ; %
        \edge[color=MidnightBlue, line width=1pt]{h2} {y1} ; %
      }
      & \hspace{1.5cm} \mbox{}\tikz{ %
        \node[\obs] (y1) {$Y_1$} ; %
        \node[\obs, right=of y1] (y2) {$Y_2$} ; %
        \node[\obs, below=of y1, yshift=0cm] (h1) {$X_1$} ; %
        \node[\obs, below=of y2, yshift=0cm] (h2) {$X_2$} ; %
        \edge[color=black, line width=1pt]{h1} {y1} ; %
        \edge[color=black, line width=1pt]{h2} {y2} ; %
        \edge[-, color=orange, line width=1.25pt]{y1}{y2} ; %
      } \\
      \hspace{.5cm}(a) \mbox{No interference } 
      & \hspace{1.5cm}(b) \mbox{Interference:} 
      & \hspace{1.5cm}(c) \mbox{Interference:}\\
      \hspace{.5cm} & \hspace{1.5cm}\mbox{spillover} & \hspace{1.5cm} \mbox{contagion}\s\s\s\\
    & \hspace{1.5cm} \mbox{} 
      \tikz{ %
        \node[\obs] (y1) {$Y_1$} ; %
        \node[\obs, right=of y1] (y2) {$Y_2$} ; %
        \node[\obs, below=of y1, yshift=0cm] (h1) {$X_1$} ; %
        \node[\obs, below=of y2, yshift=0cm] (h2) {$X_2$} ; %
        \edge[color=black, line width=1pt]{h1} {y1} ; %
        \edge[color=MidnightBlue, line width=1.15pt]{h1} {y2} ; %
        \edge[color=black, line width=1pt]{h2} {y2} ; %
        \edge[color=MidnightBlue, line width=1pt]{h2} {y1} ; %
        \edge[-, color=orange, line width=1.25pt]{y1}{y2} ; %
      }
    & \\ 
    & \hspace{1.5cm}(d) \mbox{Interference:} & \\
    & \hspace{1.5cm}{\mbox{spillover and contagion}} &\\
\end{tabular}
\caption{\label{graph.ts.os}
A graphical representation of a population with $N = 2$ units 
$1, 2$,
treatment assignments $X_1, X_2$,
and outcomes $Y_1, Y_2$,
with causal effects represented by directed lines and dependence represented by undirected lines.
The black- and green-colored lines represent causal effects of treatment assignments $X_1, X_2$ on outcomes $Y_1, Y_2$ due to treatment and spillover.
The orange-colored line between $Y_1$ and $Y_2$ represents dependence conditional on $X_1$ and $X_2$ induced by contagion;
note that when $Y_1$ and $Y_2$ are observed at a single time point,
contagion cannot be identified as a causal process:
see Figure \ref{graph.contagion}.
}
\end{figure}
\subsection{Spillover}
\label{sec:treatment.spillover}

Spillover occurs when the potential outcomes of units $i$ are directly affected by the treatment assignments of other units $j$.
In the running example,
spillover occurs when teenager $i$'s friend $j$ is exposed to advertisements of designer clothes and shares them with $i$.

\subsection{Contagion}
\label{sec:outcome.spillover}

Contagion occurs when the potential outcomes of units $i$ are indirectly affected by the treatment assignments of other units $j$ via their potential outcomes.
In the running example,
contagion can occur when teenager $i$'s friend $j$ is exposed to advertisements of designer clothes and purchases the advertised designer clothes.
While $i$ may not be aware that $j$ was exposed to advertisements of designer clothes,
$i$ may observe $j$ wearing designer clothes and may admire them,
so $i$ may purchase them as well.

\begin{figure}
\vspace{-.25cm}
\begin{center}
\begin{tabular}{ccc}
      \hspace{0.5cm}\mbox{} \tikz{ %
        \node[\obs] (y1) {$Y_1$} ; %
        \node[\obs, right=of y1] (y2) {$Y_2$} ; %
        \node[\obs, below=of y1, yshift=0cm] (h1) {$X_1$} ; %
        \node[\obs, below=of y2, yshift=0cm] (h2) {$X_2$} ; %
        \edge[color=black, line width=1pt]{h1} {y1} ; %
        \edge[color=MidnightBlue, line width=1.15pt]{h1} {y2} ; %
        \edge[color=black, line width=1pt]{h2} {y2} ; %
        \edge[color=MidnightBlue, line width=1pt]{h2} {y1} ; %
        \edge[color=orange, line width=1.25pt]{y1}{y2} ; %
      }
&
       \hspace{1.5cm}\mbox{}\tikz{ %
        \node[\obs] (y1) {$Y_1$} ; %
        \node[\obs, right=of y1] (y2) {$Y_2$} ; %
        \node[\obs, below=of y1, yshift=0cm] (h1) {$X_1$} ; %
        \node[\obs, below=of y2, yshift=0cm] (h2) {$X_2$} ; %
        \edge[color=black, line width=1pt]{h1} {y1} ; %
        \edge[color=MidnightBlue, line width=1.15pt]{h1} {y2} ; %
        \edge[color=black, line width=1pt]{h2} {y2} ; %
        \edge[color=MidnightBlue, line width=1pt]{h2} {y1} ; %
        \edge[color=orange, line width=1.25pt]{y2}{y1} ; %
      }
& 
       \hspace{1.5cm}\mbox{}\tikz{ %
        \node[\obs] (y1) {$Y_1$} ; %
        \node[\obs, right=of y1] (y2) {$Y_2$} ; %
        \node[\obs, below=of y1, yshift=0cm] (h1) {$X_1$} ; %
        \node[\obs, below=of y2, yshift=0cm] (h2) {$X_2$} ; %
        \node[\lat, above=of y1, xshift=0.85cm, yshift=-.25cm] (u) {$U$} ; %
        \edge[color=black, line width=1pt]{h1} {y1} ; %
        \edge[color=MidnightBlue, line width=1.15pt]{h1} {y2} ; %
        \edge[color=black, line width=1pt]{h2} {y2} ; %
        \edge[color=MidnightBlue, line width=1pt]{h2} {y1} ; %
        \edge[color=orange, line width=1.25pt]{u}{y1,y2} ; %
      }
\\
 \hspace{0.5cm}(a) Contagion
 &\hspace{1.5cm} (b) Contagion
&\hspace{1.5cm} (c) Unobserved $U$:
\\
&& \hspace{2.3cm} common source
\end{tabular}
\caption{\label{graph.contagion}
If $Y_1$ and $Y_2$ are observed at a single time point,
we cannot distinguish whether
(a) $Y_1$ affects $Y_2$,
(b) $Y_2$ affects $Y_1$,
or (c) the dependence between $Y_1$ and $Y_2$ conditional on $X_1$ and $X_2$ is induced by an unobserved variable $U$ that affects both $Y_1$ and $Y_2$.
As a result,
contagion cannot be identified as a causal process,
but it can be accommodated by capturing dependence between $Y_1$ and $Y_2$ conditional on $X_1$ and $X_2$.
The source of dependence in examples (a) and (b) is contagion,
whereas the source of dependence in example (c) is an unobserved variable $U$.
}
\end{center}
\end{figure}

\subsection{Contagion: Causation versus Correlation}
\label{sec:causation.versus.correlation}

If outcomes are observed at a single time point (as considered here) rather than multiple time points \citep[as considered in][]{OgVW14},
the observed outcomes can be viewed as a single realization from the stationary distribution of a time-indexed stochastic process \citep{OgShLe20,LaRi02}.
If no additional information is available,
it is impossible to distinguish whether,
for any pair of distinct units $i$ and $j$,
(a) $Y_i$ affects $Y_j$,
(b) $Y_j$ affects $Y_i$,
or (c) there is an unobserved variable that affects both $Y_i$ and $Y_j$.
A graphical representation of scenarios (a),
(b),
and (c) can be found in Figure \ref{graph.contagion}.
Here, 
(a) and (b) are instances of contagion whereas (c) is not, 
but all three scenarios can lead to statistical dependence among outcomes $Y_1, \ldots, Y_N$ conditional on treatment assignments $X_1, \ldots, X_N$.
Therefore,
without a causal ordering among outcomes,
contagion cannot be identified as a causal process.
That said,
contagion can be accommodated by capturing dependence among outcomes $Y_1, \ldots, Y_N$ conditional on treatment assignments $X_1, \ldots, X_N$, while acknowledging that such dependence may arise from other sources, 
including (c).
We first discuss how dependence induced by contagion can be captured in Section \ref{sec:capturing.contagion},
and then mention that there are other possible sources of dependence in Section \ref{sec:other.sources}.

\subsection{Capturing Contagion}
\label{sec:capturing.contagion}

While contagion cannot be identified as a causal process based on outcomes observed at a single time point, 
contagion can be accommodated  by
\bi 
\item modeling dependence of outcomes $Y_1, \ldots, Y_N$ conditional on treatment assignments $X_1, \ldots, X_N$;
\vspace{.05cm}
\item or modeling dependence among fixed potential outcomes $y_1(\bx), \ldots, y_N(\bx)$ or random potential outcomes $Y_1(\bx), \ldots, Y_N(\bx)$ ($\bx \in \mX$).
\ei 
These two approaches are equivalent under igorability assumptions (Section \ref{sec:identification}).
In both the fixed and random potential outcome frameworks, 
there have been attempts to model dependence among potential outcomes that accommodate contagion \cite{leung2022causal,OgShLe20,TTFuSh21,BhSe24}.
These attempts may be deterministic or probabilistic,
depending on whether potential outcomes are fixed or random.

\paragraph*{Fixed potential outcomes}

In the fixed potential outcome framework,
contagion can be accommodated by assuming that,
for all possible treatment assignments $\bx\in\mX$, 
the potential outcomes $y_i(\bx)$ of units $i$ are related by deterministic functions $g_i$ of the form
$$y_i(\bx) = g_i(\bx, \by_{-i}(\bx), \epsilon_i),$$
where $\epsilon_i$ are non-random errors.
In general, 
it is not guaranteed that there exist potential outcomes $\by(\bx) \coloneqq (y_1(\bx), \ldots, y_N(\bx))$ satisfying the above equations.
A simple example that does satisfy the above equations is the contagion model mentioned on page 4 of \citet{yu2022estimating},
which assumes that,
for all $\bx \in \mX$,
\beno
y_i(\bx) 
= \alpha_i + \beta_i\, x_i 
+ \dsum_{j \neq i}^N \delta_{i,j}\, y_j(\bx),
\ee
where $\alpha_i \in \mbR$,
$\beta_i \in \mbR$,
and $\delta_{i,j} \in \mbR$ may be interpreted as the intercept,
the main effect of treatment,
and the effect of contagion,
respectively.
If $\delta_{i,j} = 0$ for all $(i, j)$,
there is no contagion and the $N$-vector of potential outcomes $\by(\bx)$ reduces to
\beno
\label{contagion1}
\by(\bx) 
&=& \bm\alpha + \bm{B}\, \bx,
\ee
otherwise
\be
\label{contagion2}
\by(\bx)
&=& (\bI_N - \bm\Delta)^{-1}\, \bm{\alpha} + \dsum_{d=0}^{\infty} \bm\Delta^d \bm{B}\, \bx,
\ee
where the $N \times N$-matrix $\bm{B}$ contains elements $\beta_i$ on its main diagonal and elements $0$ off its main diagonal,
while the $N \times N$-matrix $\bm\Delta$ contains elements $0$ on its main diagonal and elements $\delta_{i,j}$ off its main diagonal.
Equation \eqref{contagion2} reveals that contagion gives rise to interference,
because it violates the no-interference assumption
\beno
y_i(x_i; \bx_{-i}) = y_i(x_i; \bx_{-i}') \text{ for all } i,\, x_i,\, \bx_{-i},\, \bx_{-i}'.
\ee

\paragraph*{Random potential outcomes}

In the random potential outcome framework, contagion can be captured by assuming that,
for all $\bx\in\mX$, 
the conditional distribution of $Y_i(\bx)$ depends on $\by_{-i}(\bx)$:
$$Y_i(\bx) \mid \{\bY_{-i}(\bx) = \by_{-i}(\bx)\}\;
=\; g_i(\bx, \by_{-i}(\bx),\epsilon_i),$$
where $\epsilon_i$ are random errors.
Similar to the fixed potential outcome framework,
the above assumption is not valid in general, 
and its validity requires the conditional distributions of $Y_i(\bx)$ to be consistent with a valid joint probability distribution of $\bY(\bx)\coloneqq (Y_1(\bx),\ldots,Y_N(\bx))$.
A simple example is a linear model of the form
\beno
\label{equation2}
&& Y_i(\bx) \mid \{\bY_{-i}(\bx) = \by_{-i}(\bx)\}
\s
\\
&=&\, \alpha_i + \beta_i\, x_i + \dsum_{j\neq i}^N \delta_{i,j}\, y_j(\bx) + \epsilon_i.
\ee
If the random errors $\epsilon_i$ are Gaussian, 
$\bY(\bx)$ is multivariate Gaussian under some restrictions \citep{rue2005gaussian}.

\subsection{Other Sources of Dependence}
\label{sec:other.sources}

A complication is that statistical dependence can arise from sources other than contagion:
e.g.,
dependence between the outcomes $Y_i$ and $Y_j$ of a pair of distinct units $i$ and $j$ conditional on treatment assignments $X_1, \ldots, X_N$ can arise from a common but unobserved source that affects both $Y_i$ and $Y_j$ \citep[e.g., unobserved similarity or homophily,][]{ShTh11}.
A graphical representation can be found in Figure \ref{graph.contagion}.
Throughout the remainder of the survey,
we review literature on interference in the form of spillover as well as contagion inducing dependence among outcomes,
and note that few publications have considered other sources of dependence.

\subsection{Representations of Interference}
\label{sec:interference.representations}

We discuss two approaches to representing interference,
based on interference graphs (Section \ref{sec:interference.graphs}) and exposure mappings (Section \ref{sec:exposure.mappings}).
These two approaches complement each other and can be combined.While some publications use neither of them \citep[e.g.,][]{choi2017estimation, savje2021average, hu2022average},
many others use one of them or both.

\subsubsection{Interference Graph}
\label{sec:interference.graphs}

{\blues To represent who can affect the potential outcomes of whom,
interference graphs can be used.
A convenient representation of an interference graph is a $N \times N$-matrix $\bZ \in \{0,\, 1\}^{N \times N}$ with elements $Z_{i,j}$,
where $Z_{i,j} \coloneqq 1$ indicates that $i$ and $j$ are neighbors in the interference graph and $Z_{i,j} \coloneqq 0$ otherwise.The interference graph can be used to capture interference: 
e.g.,
the potential outcomes of unit $i$ may be affected by the treatment assignments of its neighbors $j$ in the interference graph,
which is known as neighborhood exposure.
Some recent publications relax neighborhood exposure assumptions,
by permitting interference by both neighbors and non-neighbors who are connected by a path in the interference graph \citep[e.g.,][]{TTFuSh21,leung2022causal, belloni2025neighborhood, leung2026cluster}. 
We discuss these weaker assumptions in Section \ref{assumption1}.

Throughout the remainder of the survey,
we consider undirected connections and exclude self-connections,
so that $Z_{i,i} \coloneqq 0$ and $Z_{i,j} = Z_{j,i}$.}
In a mild abuse of language,
we refer to $\bZ$ as the interference graph.
In principle, 
the interference graph can have weighted connections,
where the weights quantify the number of interactions or the strength of the connections among units.
Having said that,
we consider binary interference graphs $\bZ \in \mZ \coloneqq \{0, 1\}^{N \times N}$ throughout the survey, 
with the exception of the weighted interference graphs in Section \ref{sec:spatial.interference} capturing spatial interference.

\subsubsection{Exposure Mappings}
\label{sec:exposure.mappings}

Exposure mappings help represent how the potential outcomes of units $i$ are affected by the treatment assignments of other units. 
Exposure mappings are functions $f_i: \mX \mapsto \mcF_i$ from the set of all possible treatment assignments $\mX$ of size $|\mX| \leq 2^N$ to sets $\mcF_i$ of sizes $|\mcF_i| \ll |\mX|$ \citep{aronow2017estimating}.
An exposure mapping $f_i$ specifies how the potential outcomes of unit $i$ are affected by the treatment assignments of others,
in the sense that
\beno
f_i(\bx) = f_i(\bx')
\text{ implies } y_i(\bx) = y_i(\bx')
\ee
when potential outcomes are fixed,
and
\beno
f_i(\bx) = f_i(\bx')
\text{ implies } Y_i(\bx) = Y_i(\bx')
\ee
when potential outcomes are random.
Exposure mappings can depend on a given interference graph $\bz$ \citep[e.g.,][]{ugander2013graph,SuAi17, ogburn2024causal, jagadeesan2020designs,leung2020treatment,LiWa22, ogburn2024causal}.
An example is
\beno
f_i(\bx)
&=& (x_i,\, \sum_{j=1}^N x_j\, z_{i,j},\, \sum_{j=1}^N z_{i,j}).
\ee

\section{Targets of Causal Inference}
\label{sec:target}

The primary target of causal inference is the causal effect of treatments on outcomes.

If there is no interference and the potential outcomes $y_i(1)$ and $y_i(0)$ of units $i$ are fixed,
differences in potential outcomes $y_i(1) - y_i(0)$ can be attributed to the causal effect of treatments on outcomes.
In accordance,
the average treatment effect (ATE) is defined by 
\beno
\tau 
&\coloneqq& \dfrac{1}{N} \dsum_{i=1}^N \,(y_i(1) - y_i(0)).
\ee

In the presence of interference,
there are no universal definitions of causal effects.
We first review selected causal effects based on fixed potential outcomes and interference graphs in Section \ref{sec:causal.effects.fixed},
and causal effects based on random potential outcomes and interference graphs in Section \ref{sec:causal.effects.random}.
We then discuss identification of causal effects in Section \ref{sec:identification}.
Throughout Section \ref{sec:target},
we assume that there is an underlying interference graph.
For ease of presentation,
we consider the interference graph to be fixed when potential outcomes are fixed,
and random when potential outcomes are random,
although some publications combine random potential outcomes with fixed interference graphs \citep[e.g.,][]{ogburn2024causal}.
If the interference graph is random,
causal effects may condition on it.

\subsection{Causal Effects: Fixed Potential Outcomes}
\label{sec:causal.effects.fixed}

We first introduce causal effects based on fixed potential outcomes and interference graph $(\by(\cdot), \bz)$.

For any combination of treatment assignments $\bx \in \mX$,
define the average potential outcome by
\beno
\mu(\bx)
\;\coloneqq\; \dfrac{1}{N} \dsum_{i=1}^N y_i(\bx).
\ee 
Then the effect of treatment assignments $\bx_1$ relative to treatment assignments $\bx_2 \neq \bx_1$ can be quantified by
\beno
\tau(\bx_1,\bx_2)
&\coloneqq& \mu(\bx_1) - \mu(\bx_2).
\ee
An example,
known as global average treatment effect (GATE) \citep{ugander2013graph} and total treatment effect (TTE) \citep{yu2022estimating},
is
\beno
\tau(\bm{1}_N, \bm{0}_N)
&\coloneqq& \mu(\bm{1}_N) - \mu(\bm{0}_N).
\eeThe causal effect $\tau(\bm{1}_N, \bm{0}_N)$ compares potential outcomes in two scenarios:
the scenario in which no unit is treated ($\bm{0}_N$),
and the scenario in which all units are treated ($\bm{1}_N$).
For example,
companies,
governments,
or other organizations may be interested in assigning a treatment to all eligible population members and may therefore wish to compare $\mu(\bm{1}_N)$ to $\mu(\bm{0}_N)$.
In other scenarios, decision-makers may wish to compare the effect of two intermediate treatment assignments $\bx_1$ and $\bx_2 \neq \bx_1$ using $\tau(\bx_1, \bx_2)$.
In the running example,
advertisers may be interested in intermediate treatment regimes in which some but not all teenagers are targeted with advertisements of designer clothes:
targeting all teenagers may not be cost-effective,
because teenagers may purchase designer clothes without being exposed to advertisements,
by virtue of spillover and contagion.

Other causal effects are based on exposure mappings.
Let the average potential outcome at exposure level $d$ be
\beno
\mu^E(d)
&\coloneqq& \dfrac{1}{N} \dsum_{i=1}^N y_i^E(d),
\ee 
where $y_i^E(d)$ is the potential outcome of unit $i$ when $f_i(\bx)=d$.
The effect of exposure to level $d_1$ relative to level $d_2 \neq d_1$ can be quantified by \citep{aronow2017estimating}
\beno
\label{diff.effect.1}
\tau^E(d_1, d_2)
&\coloneqq& \mu^E(d_1) - \mu^E(d_2).
\ee

If investigators wish to average over all possible treatment assignments $\bx \in \mX$ rather than comparing two specified treatment assignments $\bx_1$ and $\bx_2$ or exposure levels $d_1$ and $d_2$,
the following causal effect can be used \cite{hu2022average}:
\beno
\tau_{\mX}
&\coloneqq& \dfrac{1}{N} \dsum_{i=1}^N \dsum_{j =1}^N \mbE_{\mX} [y_j(x_i = 1; \bX_{-i})-y_j(x_i = 0; \bX_{-i})].
\ee
The causal effect $\tau_{\mX}$ can be decomposed into direct and indirect causal effects:
\beno
\tau^D_{\mX}
&\coloneqq& \dfrac{1}{N} \dsum_{i=1}^N \mbE_{\mX} [y_i(x_i = 1; \bX_{-i})-y_i(x_i = 0; \bX_{-i})]\vspace{.05cm}
\\
\tau^I_{\mX}
&\coloneqq& \dfrac{1}{N} \dsum_{i=1}^N \dsum_{j \neq i}^N \mbE_{\mX} [y_j(x_i = 1; \bX_{-i})-y_j(x_i = 0; \bX_{-i})].
\ee
The direct and indirect causal effects $\tau^D_{\mX}$ and $\tau^I_{\mX}$ can be interpreted as follows:
\bi
\item The direct causal effect $\tau^D_{\mX}$ quantifies the effect of unit $i$'s treatment assignment $x_i$ on its potential outcome $y_i(x_i; \bx_{-i})$,
averaged over all $i$.
\vspace{.05cm}
\item The indirect causal effect $\tau^I_{\mX}$ quantifies the effect of unit $i$'s treatment assignment $x_i$ on the potential outcomes $y_j(x_i; \bx_{-i})$ of other units $j$,
averaged over all $i \neq j$.
\ei 
{\blues In the special case of $X_i\, \ind\, \mbox{Bernoulli}(\pi_i)$ ($\pi_i \in (0, 1)$),
the causal effect $\tau_{\mX}$ can be expressed as
\beno 
\tau_{\mX} 
&=& \dfrac{1}{N} \dsum_{i=1}^N \dfrac{\partial}{\partial\, \pi_i}\, \mbE_{\mX}\, Y_i + \dfrac{1}{N} \dsum_{i=1}^N \dsum_{j \neq i}^N \dfrac{\partial}{\partial\, \pi_j}\, \mbE_{\mX}\, Y_i,
\ee 
where $Y_i = y_i(\bX)$ by the consistency assumption in Equation \eqref{consistency.assumption}.
If the vector of treatment assignment probabilities $(\pi_1, \ldots, \pi_N)$ is interpreted as a policy,
then $\tau_{\mX}$ quantifies how sensitive expected outcomes are to changes in the policy $(\pi_1, \ldots, \pi_N)$ \citep{hu2022average}.}

Direct and indirect causal effects based on exposure mappings can be defined as well \citep[see][]{forastiere2021identification}.

\subsection{Causal Effects: Random Potential Outcomes}
\label{sec:causal.effects.random}

A related class of causal effects is based on random potential outcomes and interference graphs $(\bY(\cdot), \bZ)$.

A natural extension of the mean potential outcome $\mu(\bx)$ from fixed to random potential outcomes is
\beno
\mu_{\mY(\cdot)\mid\mZ}(\bx)
&\coloneqq& \dfrac{1}{N} \dsum_{i=1}^N \mbE_{\mY(\cdot)\mid\mZ}\, [Y_i(\bx)\mid \bZ=\bz].
\ee
The mean $\mu_{\mY(\cdot)\mid\mZ}(\bx)$ averages over all possible values of potential outcomes $\bY(\bx)$ conditional on a known interference graph $\bZ = \bz$.
To average over all possible interference graphs $\bz \in \mZ$ rather than conditioning on a known interference graph,
one may use
\beno
\mu_{\mY(\cdot),\mZ}(\bx)
&\coloneqq& \dfrac{1}{N} \dsum_{i=1}^N \mbE_{\mY(\cdot),\mZ}\, Y_i(\bx).
\ee
A relative of $\mu_{\mY(\cdot),\mZ}(\bx)$ based on exposure mappings is
\beno
\mu_{\mY(\cdot),\mZ}^E(d)
&\coloneqq& \dfrac{1}{N} \dsum_{i=1}^N \mbE_{\mY(\cdot),\mZ}\, Y_i^E(d).
\ee
The effects of treatment assignments $\bx_1$ and $\bx_2$ or exposure levels $d_1$ and $d_2$ can be assessed based on 
$\mu_{\mY(\cdot),\mZ}(\bx_1) - \mu_{\mY(\cdot),\mZ}(\bx_2)$
or
$\mu_{\mY(\cdot),\mZ}^E(d_1) - \mu_{\mY(\cdot),\mZ}^E(d_2)$.
\vspace{.1cm}

The causal effect $\tau_{\mX}$ can be extended from fixed to random potential outcomes and interference graphs \cite{LiWa22}:
\beno
\tau_{\mX,\mY(\cdot),\mZ}
&\coloneqq& \dfrac{1}{N} \dsum_{i=1}^N \dsum_{j=1}^N \mbE_{\mX,\mY(\cdot),\mZ}\, [Y_j(x_i=1; \bX_{-i})-  Y_j(x_i=0; \bX_{-i})].
\ee
The causal effect $\tau_{\mX,\mY(\cdot),\mZ}$ averages over all possible treatment assignments,
potential outcomes, and interference graphs.
Since causal effects based on fixed potential outcomes and interference graphs $(\by(\cdot), \bz)$ can be interpreted as conditional causal effects given $(\bY(\cdot), \bZ) = (\by(\cdot), \bz)$,\,
$\tau_{\mX,\mY(\cdot),\mZ}$ is related to $\tau_\mX$ as follows:
\beno 
\tau_{\mX,\mY(\cdot),\mZ} 
&=& \mbE_{\mY(\cdot),\mZ}\, \tau_\mX.
\ee
In accordance with $\tau_{\mX}$,
$\tau_{\mX,\mY(\cdot),\mZ}$ can be decomposed into direct and indirect causal effects $\tau^D_{\mX,\mY(\cdot),\mZ}$ and $\tau^I_{\mX,\mY(\cdot),\mZ}$.

It is worth noting that going from $\mu(\bx)$ to $\mu_{\mY(\cdot),\mZ}(\bx)$,
and from $\tau_{\mX}$ to $\tau_{\mX,\mY(\cdot),\mZ}$,
may require additional assumptions about the probability law generating $(\bX, \bY(\cdot), \bZ)$.
We discuss in Section \ref{subsec:random} when it makes sense to make such additional assumptions.

\subsection{Identification of Causal Effects}
\label{sec:identification}

The Neyman-Rubin framework assesses causal effects of treatments on potential outcomes by comparing two or more sets of potential outcomes.
The fundamental problem of causal inference in the Neyman-Rubin framework is that a single set of potential outcomes is observed,
whereas all other $2^N - 1$ sets of potential outcomes are missing.
This raises a critical question:
under which conditions can causal effects based on potential outcomes be identified based on observed outcomes?

In the absence of interference,
\citet{hernan2020causal} provide a comprehensive review of identification.
A causal effect is said to be nonparametrically identified if it can be computed from the distribution of the observed data:
that is, 
the distribution of the observed data is compatible with one and only one value of the causal effect.
Without interference, 
identification is guaranteed under three conditions: 
consistency defined in Equation \eqref{consistency.assumption},
ignorability or conditional ignorability,
and positivity,
defined below.
Under interference, 
these conditions alone may not guarantee identification.
The primary challenge is the lack of independent replications under interference,
as discussed by \citet{TTFuSh21} and others.
Some works, 
including \citet{leung2020treatment} and \citet{ogburn2024causal}, 
consider nonparametric structural equation models for outcomes with independent errors, 
under which identification can be established.
A full-fledged discussion of identification is beyond the scope of the survey,
because identification under interference is a challenging topic in its own right and we cover a wide range of causal effects.
That said,
we review ignorability,
conditional ignorability,
and positivity,
which help relate the mean of potential outcomes $\bY(\cdot)$ to the mean of observed outcomes $\bY$ and are necessary but may not be sufficient for nonparametric identification under interference.
To state these assumptions,
we assume that treatment assignments $\bX$,
potential outcomes $\bY(\cdot)$,
and interference graph $\bZ$ are random,
without making assumptions about the probability law of $(\bX, \bY(\cdot), \bZ)$.

Ignorability is satisfied in experimental settings in which investigators have complete control over the treatment assignment mechanism and ensure that treatment assignments $\bX$ are independent of potential outcomes $\{\bY(\bx):\, \bx \in \mX\}$ given interference graph $\bZ = \bz$:

\begin{condition}
\label{con:ignorability1}
{\bf Ignorability.}\;
For all $\bz \in \mZ$,
\be
\label{ignorability1}
\bX
\;&\orth&\;
\{\bY(\bx):\, \bx \in \mX\} \mid \bZ = \bz.\\
\ee
\end{condition}

\vspace{-.55cm}

\noindent
Condition \ref{con:ignorability1} allows treatment assignments $\bX$ to depend on interference graph $\bZ = \bz$,
which helps reduce the variance of estimators under interference (see Section \ref{sec:variance.reduction}).

Outside of experimental settings,
ignorability may hold conditional on observed confounders $\bC\in\mC$ \citep{TTFuSh21,ogburn2024causal,BhSe24}:

\vspace{-.05cm}

\begin{condition}
\label{con:ignorability2}
{\bf Conditional Ignorability.}\;
For all $(\bc, \bz) \in \mC \times \mZ$,
\beno
\bX \orth \{\bY(\bx):\, \bx \in \mX\} \mid (\bC, \bZ) = (\bc, \bz).
\ee
\end{condition}
\vspace{-.75cm}

Many causal effects are averages of differences between potential outcomes at different treatment combinations.
A positivity assumption is required to ensure that the potential outcomes $\bY(\bx)$ are observed with positive probability under all treatment assignments $\bx \in \mX \subseteq \{0, 1\}^N$.
A strong positivity condition,
which suffices for all average causal effects but can be weakened for specific causal effects,
is the following:
\begin{condition}
    \label{con:positivity}
    {\bf Positivity.}\; For all $\bz\in\mZ$,
    \beno
    \mbP_{\mX\mid\mZ}(\bX=\bx\mid\bZ=\bz) > 0 \ \text{for all }\bx\in\{0,1\}^N.
    \ee
\end{condition}

\vspace{-.75cm}

To demonstrate that ignorability and positivity help relate potential outcomes $\bY(\cdot)$ to observed outcomes $\bY \mid \bX = \bx$,
we focus on experimental settings \cite{jagadeesan2020designs,leung2020treatment}.
In these settings,
the ignorability assumption in Equation \eqref{ignorability1} along with the consistency assumption in Equation \eqref{consistency.assumption} imply that the distribution of potential outcomes $\bY(\bx) \mid \bZ = \bz$ is identical to the distribution of observed outcomes $\bY \mid (\bX, \bZ) = (\bx, \bz)$:
\be
\label{YxtoY}
\bY(\bx) \mid \bZ = \bz
\;\;&\stackrel{d}{=}&\;\; \bY \mid (\bX, \bZ) = (\bx, \bz).
\ee
Equations \eqref{ignorability1} and \eqref{YxtoY} help express the mean $\mu_{\mY(\cdot)\mid\mZ}(\bx)$ of potential outcomes $Y_i(\bx)$ as the conditional mean of observed outcomes $Y_i$:
\beno
\mu_{\mY(\cdot)\mid\mZ}(\bx)
&\coloneqq& \dfrac{1}{N} \dsum_{i=1}^N \mbE_{\mY(\cdot)\mid\mZ}\, [Y_i(\bx)\mid \bZ = \bz] &=& \dfrac{1}{N} \dsum_{i=1}^N \mbE_{\mY \mid \mX,\mZ}[Y_i\mid (\bX, \bZ) = (\bx, \bz)].
\ee
By an argument along the same lines,
\beno
\mu_{\mY(\cdot),\mZ}(\bx)
&\coloneqq& \dfrac{1}{N} \dsum_{i=1}^N \mbE_{\mY(\cdot),\mZ}\, Y_i(\bx)
\ee
can be expressed as
\beno
&& \dfrac{1}{N} \dsum_{i=1}^N \mbE_{\mY(\cdot),\mZ}\, Y_i(\bx)\vspace{.175cm}
\\
&=&\dfrac{1}{N} \dsum_{i=1}^N \dsum_{\bz\in\mZ}\,  \mbE_{\mY(\cdot) \mid \mZ}[Y_i(\bx)\mid \bZ=\bz]\, \mbP_{\mZ}(\bZ = \bz)\vspace{.1cm}
\\
&=& \dfrac{1}{N} \dsum_{i=1}^N \dsum_{\bz\in\mZ}\, \mbE_{\mY \mid \mX,\mZ}[Y_i\mid (\bX, \bZ) = (\bx, \bz)]\, \mbP_{\mZ}(\bZ = \bz).
\ee

Weaker forms of ignorability may suffice,
depending on the causal effect of interest.
For example,
consider $\mu^E(d)$, 
the mean potential outcome at exposure level $d$.
Then the following form of ignorability is sufficient for expressing the mean of potential outcomes $\bY(\cdot)$ as the mean of observed outcomes $\bY$
\citep{forastiere2021identification,bong2025heterogeneous,khatami2025gml}:
\be
\label{ignorability2}
f_i(\bX)
\;&\orth&\;
\{Y_i(\bx):\, \bx \in \mX\} \mid (\bC,\bZ) = (\bc,\bz),
\ee
where $f_i(\bX)$ is the exposure of unit $i$. 
Two examples of $f_i(\bX)$ are
\beno                                     
f_i(\bX)                                         
&=& (X_i,\, \{X_j: z_{i,j}=1\}),                      
\ee
which is used in \citet{bong2025heterogeneous},
and
\beno                                     
f_i(\bX)                                         
&=& (X_i,\, \sum_{j=1}^N X_j\, z_{i,j},\, \sum_{j=1}^N z_{i,j}).                               
\ee
Similar conditions are used by \citet{forastiere2021identification} and \citet{khatami2025gml}.In the absence of confounders $\bC$, 
Equation \eqref{ignorability2} implies that
\beno
\mu_{\mY(\cdot),\mZ}^E(d)
&\coloneqq& \dfrac{1}{N} \dsum_{i=1}^N \mbE_{\mY(\cdot),\mZ}\, Y_i^E(d)
\ee
can be expressed as 
\beno
&& \dfrac{1}{N} \dsum_{i=1}^N \mbE_{\mY_i(\cdot),\mZ}\, Y_i^E(d)\vspace{.1cm}
\\
&=& \dfrac{1}{N} \dsum_{i=1}^N \dsum_{\bz \in \mZ} \mbE_{\mY_i(\cdot)\mid\mZ}[Y_i^E(d) \mid \bZ = \bz]\, \mbP_{\mZ}(\bZ = \bz)\vspace{.1cm}
\\
&=& \dfrac{1}{N} \dsum_{i=1}^N \dsum_{\bz \in \mZ} \mbE_{\mY_i\mid\mcF_i,\mZ}[Y_i \mid (F_i, \bZ) = (d, \bz)]\, \mbP_{\mZ}(\bZ = \bz),
\ee
where $F_i\coloneqq f_i(\bX) \in \mathscr{F}_i$.

Finally,
we demonstrate that the causal effect $\tau_{\mX,\mY(\cdot),\mZ}$ based on potential outcomes $\bY(\cdot)$ can be related to observed outcomes $\bY$.
Consider any pair of units $(i,j)$ and any treatment assignment $x_i \in \{0, 1\}$,
and note that
\beno
\label{identification}
&&\mbE_{\mX,\mY(\cdot),\mZ}\, Y_j(x_i; \bX_{-i})\s
\\
&=& \mbE_{\mathscr{X},\mathscr{Z}}\, \mbE_{\mathscr{Y}(\cdot) \mid \mathscr{X}, \mathscr{Z}}[Y_j(x_i; \bX_{-i}) \mid (\bX, \bZ)]\s
\\
&=& \mbE_{\mathscr{X},\mathscr{Z}}\, \mbE_{\mathscr{Y}(\cdot) \mid \mathscr{Z}}[Y_j(x_i; \bX_{-i}) \mid \bZ]\s
\\
&=& \mbE_{\mathscr{X},\mathscr{Z}}\, \mbE_{\mathscr{Y} \mid \mathscr{\bX}, \mathscr{Z}}[Y_j \mid (X_i=x_i, \bX_{-i}, \bZ)],
\ee
where the second identity follows from Equation \eqref{ignorability1} while the third identity follows from Equation \eqref{YxtoY}.
We can therefore express the causal effect $\tau_{\mX,\mY(\cdot),\mZ}$ as
\beno 
&& \dfrac{1}{N} \dsum_{i=1}^N \dsum_{j=1}^N \mbE_{\mathscr{X},\mathscr{Z}}\, \mbE_{\mathscr{Y} \mid \mathscr{\bX}, \mathscr{Z}}[Y_j \mid (X_i=1, \bX_{-i}, \bZ)]\vspace{.1cm}
\\
&-& \dfrac{1}{N} \dsum_{i=1}^N \dsum_{j=1}^N \mbE_{\mathscr{X},\mathscr{Z}}\, \mbE_{\mathscr{Y} \mid \mathscr{\bX}, \mathscr{Z}}[Y_j \mid (X_i = 0, \bX_{-i}, \bZ)].
\ee

Throughout the remainder of the survey,
we assume that either ignorability or conditional ignorability together with positivity is satisfied.

Last,
but not least,
identification of causal effects is a minimum requirement:
additional assumptions may be required to estimate the mentioned regression functions (i.e., 
the conditional expectations of observed outcomes).
We review such assumptions in Section \ref{sec:comparing.design.model.based.estimators}.

\hide{

\subsection{Example: Closed-Form Expressions}
\label{sec:characterizing.causal.effects}

We introduce an example to provide insight into causal effects based on closed-form expressions,
which demonstrates that the contributions of treatment,
spillover,
and contagion to causal effects are intertwined even in the simplest possible scenarios.
\hide{
We introduce an example to provide insight into the direct, 
indirect, 
and total causal effects $\tau_{\mX}^D$,
$\tau_{\mX}^I$,
and $\tau_{\mX}$ introduced by \citet{hu2022average},
using closed-form expressions.
As acknowledged by \citet{hu2022average},
the definitions of these causal effects are abstract,
so the authors introduce simple examples of parametric models under which the direct, 
indirect, 
and total causal effects reduce to expressions interpretable as direct, 
indirect, 
and total causal effects.
These examples of accommodate spillover,
but not contagion.
We discuss closed-form expressions of the direct, 
indirect, 
and total causal effects under simple parametric model assumptions resembling those of \citet{hu2022average},
but we accommodate spillover along with correlations induced by contagion.
We then demonstrate that the contributions of treatment,
spillover, 
and contagion to these causal effects are intertwined even in the simplest possible scenarios.}

We consider two sets of assumptions. 
First, 
we assume that,
for all treatment assignments $\bx \in \mX$,
the joint probability law of random potential outcomes $\bY(\bx)$ conditional on interference graph $\bZ=\bz$ satisfies
\beno
\label{natural}
\mbE_{\mY_i(\bx)\mid \mY_{-i}(\bx),\mZ}[Y_i(\bx) \mid (\bY_{-i}(\bx), \bZ) = (\by_{-i}(\bx), \bz)]\vspace{.1cm}
\\
\hspace{-.05cm} \coloneqq \hspace{-.05cm} \beta\, x_i 
\hspace{-.05cm} + \hspace{-.05cm} \gamma\, c_{N,i,1}(\bz) \hspace{-.0675cm} \dsum_{j=1}^N x_j\, z_{i,j}\vspace{.1cm}
\hspace{-.05cm} + \hspace{-.05cm} \delta\, c_{N,i,2}(\bz) \hspace{-.0675cm} \dsum_{j=1}^N y_j(\bx)\, z_{i,j}.
\ee
The weights $\beta \in \mR$,
$\gamma \in \mR$,
and $\delta \in \mR$ may be interpreted as the main effect of treatment,
spillover,
and contagion on the expected potential outcome $Y_i(\bx)$ of unit $i$ given the potential outcomes $\bY_{-i}(\bx) = \by_{-i}(\bx)$ of all other units and interference graph $\bz$.
The quantities $c_{N,i,1}(\bz)$ and $c_{N,i,2}(\bz)$ are scaling constants that may depend on the number of units $N$,\,
unit $i$,\,
and interference graph $\bz$.
\hide{
The assumption of linear conditional expectations is satisfied by,
e.g.,
Gaussian Markov random fields \citep{rue2005gaussian,TTFuSh21},
which are some of the simplest models capturing spillover along with dependence among outcomes induced by contagion \citep{OgShLe20,TTFuSh21};
note that the assumption of linear conditional expectations is different from linear-in-means models \citep{bramoulle2009identification}:
the residuals implied by the conditional expectations stated above can be dependent due to contagion, and the covariance matrix of residuals can be a function of $\bz$ and $\delta$ \citep{BhSc25},
whereas linear-in-means models assume that residuals are independent.
}

Second, 
we assume that the interference graph $\bZ$ is generated by a low-rank model with $Z_{i,j}\, \ind\, \text{Bernoulli}(P_{i,j})$ and $\mbE(\bZ) = \bm{P} = \rho\; \bm{\alpha}\, \bm{\alpha}^\top$ \citep{latent.space.models.theory}.
The parameter $\rho \in [0, 1]$ is a sparsity parameter,
whereas $\bm{\alpha} \coloneqq (\bm{\alpha}_1, \ldots, \bm{\alpha}_N)^\top \in \mbR^{N \times K}$ is a matrix of rank $K \geq 1$ satisfying $\bm{\alpha}_i^\top \bm{\alpha}_j \in [0, 1]$.
The sparsity parameter $\rho$ may depend on $N$ and controls the sparsity of interference graph $\bZ$,
i.e.,
the expected number of connections $\mbE\, \sum_{i<j}^N Z_{i,j}$.
The vectors $\bm{\alpha}_i \in \mR^K$ and $\bm{\alpha}_j \in \mR^K$ capture unobserved heterogeneity in the propensities of units $i$ and $j$ to be connected.
Low-rank models \citep{latent.space.models.theory} are popular in statistical network analysis and include stochastic block models \citep{GaMa21} and degree-corrected block models \citep{zhao2012}.
Define
\beno
a(\delta)
&\coloneqq& \delta \lim\limits_{N\rightarrow\infty} \dfrac{1}{N}\; a_N(\delta),
\ee
where
\beno
a_N(\delta)
&\coloneqq& (\bm{1}^\top\, \bm{\alpha}) \left(\mnorm{\bm{\alpha}}_2^2\, \bI_{K} - \delta\; \bm{\alpha}^\top\bm{\alpha}\right)^{-1}(\bm{\alpha}^\top \bm{1}).
\ee

\begin{theorem}
\label{theorem:char}
Assume that $|\delta| < 1$, $c_{N,i,1}(\bZ) \coloneqq 1/\sum_{j=1}^N Z_{i,j}$ and $c_{N,i,2}(\bZ) \coloneqq 1/\mnorm{\bZ}_2$.
If $N \to \infty$,
\beno
\tau_{\mX,\mY(\cdot),\mZ}^D
\;\to\; \beta
\vspace{.15cm}
\\
\tau_{\mX,\mY(\cdot),\mZ}^I
\;\to\; \gamma + (\beta + \gamma)\, a(\delta)
\vspace{.15cm}
\\
\tau_{\mX,\mY(\cdot),\mZ}
\;\to\; (\beta + \gamma)\, (1 + a(\delta)),
\ee
where the causal effects $\tau_{\mX,\mY(\cdot),\mZ}^D$,
$\tau_{\mX,\mY(\cdot),\mZ}^I$,
and $\tau_{\mX,\mY(\cdot),\mZ}$ are defined in Section \ref{sec:causal.effects.random}.
\end{theorem}

Theorem \ref{theorem:char} is based on Corollary 1 of \citet{BhSc25}.
We first discuss the assumptions and scope of Theorem \ref{theorem:char}:
\bi 
\item {\em Treatment assignments $\bX$:}
Theorem \ref{theorem:char} does not make assumptions about the process generating treatment assignments $\bX$ and therefore covers independent and dependent treatment assignments in experimental and observational settings.
\vspace{.05cm}
\item {\em Interference graph $\bZ$:}
While Theorem \ref{theorem:char} assumes that the interference graph $\bZ$ is generated by a low-rank model,
there are more general results that do not require assumptions about the interference graph $\bZ$ \citep[e.g.,][Theorems 1 and 2]{BhSc25}.
\vspace{.05cm}
\item {\em Potential outcomes $\bY(\bx) \mid \bZ = \bz$:}
Theorem \ref{theorem:char} assumes that the conditional expectations of\break 
$Y_i(\bx) \mid (\bY_{-i}(\bx), \bZ) = (\by_{-i}(\bx), \bz)$ are linear, 
but does not impose a specific distribution on\break 
$Y_i(\bx) \mid (\bY_{-i}(\bx), \bZ) = (\by_{-i}(\bx), \bz)$.
An example satisfying such assumptions is a Gaussian Markov random field capturing spillover along with correlations among outcomes induced by contagion.
\ei 
Theorem \ref{theorem:char} implies that
\bi 
\item the effects of treatment,
spillover,
and contagion are intertwined;
\vspace{.05cm}
\item contagion can decrease or increase the causal effect $\tau_{\mX,\mY(\cdot),\mZ}$,
but it cannot cancel it or change its sign,
because $a(\delta)$ is strictly increasing on $(-1, +1)$ and satisfies $a(\delta) > -1$ and $a(0) = 0$.
\ei 
These results suggest that the effects of treatment,
spillover,
and contagion are intertwined even in the simplest possible cases,
e.g., 
the multivariate Gaussian case.
Based on these results and more general results \citep[e.g.,][Theorems 1 and 2]{BhSc25},
it is reasonable to conjecture that these effects are intertwined in general,
demonstrating that contagion and other sources of statistical dependence among outcomes complicate the interpretation of causal effects.

}

\section{Data}
\label{sec:data}

Data come in countless forms and shapes,
but it is useful to distinguish two broad classes of data,
which have implications in terms of causal inference.
These two classes of data depend on whether investigators have complete control over treatment assignments:
\bi
\item {\em Experimental Settings:}
In experimental settings with complete control over treatment assignments,
investigators can ensure that treatment assignments are independent of potential outcomes,
which ensures that ignorability is satisfied.
\vspace{.05cm}
\item {\em Observational Settings:}
In settings without complete control over treatment assignments,
investigators cannot guarantee that treatment assignments are not affected by confounders.
\ei
In line with the literature on causal inference \citep[e.g.,][p.~5]{rosenbaum2020design},
the treatment assignments $\bX$ are considered to be random in both experimental and observational settings.
The difference between experimental and observational settings is that investigators control how units are assigned to treatments in experimental settings,
but lack such control in observational settings.
We discuss these settings along with implications in terms of causal inference in Sections \ref{sec:experimental.data} and \ref{sec:observations.data},
respectively.

\subsection{Experimental Settings}
\label{sec:experimental.data}

In experimental settings with complete control over treatment assignments,
investigators can assign units to treatments at random.
Two of the simplest randomized treatment assignment mechanisms are:
\bi
\item {\em Bernoulli randomization design:} 
Investigators assign each unit $i$ to the treatment group with a known assignment probability $\pi_i$:
\beno
X_i 
&\ind& \text{Bernoulli}(\pi_i),\;
& \pi_i \in (0, 1).
\ee
\item {\em Complete randomization design:}
Investigators sample $n < N$ units at random and assign them to the treatment group.
\ei 
Both randomization designs ensure that treatment assignments $\bX$ are independent of potential outcomes $\{\bY(\bx):\, \bx \in \mX\}$,
so that ignorability is satisfied (Section \ref{sec:identification}).

In addition,
investigators can design treatment assignment mechanisms to achieve specific objectives,
including but not limited to:
\bi 
\item reducing the variance of design-based estimators for causal effects;
\vspace{.05cm}
\item maximizing effects of interventions.
\ei
We discuss these objectives in Sections \ref{sec:variance.reduction} and \ref{sec:optimal.interventions},
respectively.

\subsection{Observational Settings}
\label{sec:observations.data}

If investigators do not have complete control over treatment assignments,
treatment assignments can be affected by observed or unobserved confounders.

If all confounders $\bC$ are observed and treatment assignments $\bX$ satisfy
\beno
\bX \orth \{\bY(\bx):\, \bx \in \mX\} \mid (\bC, \bZ) = (\bc, \bz),
\ee
conditional ignorability is satisfied (Section \ref{sec:identification}),
and causal inference can be performed.

Unobserved confounders are more challenging.
An interesting case study with unobserved confounders can be found in \citet{ChBaKuMaPe22},
who leverage replications in the form of temporal observations to study the effect of conflict on forest loss in Colombia in the presence of unobserved confounders.
In the absence of temporal observations,
causal inference with unobserved confounders can rely on replications in the form of non-overlapping subpopulations \citep[i.e., partial interference:][]{HuHa08}.

\section{Statistical Tests}
\label{sec:testing}

An important area of research in causal inference under interference is the development of statistical tests.
We review recent advances in statistical testing,
including
\bi 
\item randomization tests for interference (Section \ref{sec:conditional.randomization.test.interference});
\vspace{.05cm}
\item tests for average causal effects (Section \ref{sec:statistical.tests.causal.effects}).
\ei 
Since the bulk of the literature assumes that potential outcomes and interference graphs are fixed,
we focus on fixed potential outcomes and interference graphs $(\by(\cdot), \bz)$ throughout Section \ref{sec:testing}.

\subsection{Randomization Tests for Interference}
\label{sec:conditional.randomization.test.interference}

There is a growing body of work on conditional randomization tests for interference in experimental settings.
We first review tests for detecting interference (Section \ref{sec:detect.interference}) and then discuss tests for detecting specific forms of interference (Section \ref{sec:specific.interference}).

\subsubsection{Detecting Interference}
\label{sec:detect.interference}

To detect interference when potential outcomes $\by(\cdot)$ are fixed,
it is natural to test the null hypothesis that the potential outcomes $y_i(\bx)$ of units $i$ are unaffected by the treatment assignments $x_j$ of other units $j$,
\beno
\label{test.interference}
H_0^A:\, y_i(x_i; \bx_{-i}) = y_i(x_i; \bx_{-i}') \text{ for all } i,\, x_i,\, \bx_{-i},\, \bx_{-i}',
\ee
against the alternative hypothesis
\beno
H_1^A:\, y_i(x_i; \bx_{-i}) \neq y_i(x_i; \bx_{-i}') \text{ for some } i,\, x_i,\, \bx_{-i},\, \bx_{-i}'.
\eeThe null hypothesis $H_0^A$ is not sharp,
because it does not specify all potential outcomes \citep{athey2018exact}.
To demonstrate that $H_0^A$ is not sharp,
assume that $H_0^A$ is true.
Then each unit $i$ has two distinct potential outcomes $y_i(0; \bm{0}_{N-1})$ and $y_i(1; \bm{0}_{N-1})$,
noting that 
\beno 
y_i(0; \bx_{N-1})
= y_i(0; \bm{0}_{N-1})
\mbox{ for all } \bx_{N-1} \in \{0,1\}^{N-1}
\s 
\\
y_i(1; \bx_{N-1})
= y_i(1; \bm{0}_{N-1}) 
\mbox{ for all } \bx_{N-1} \in \{0,1\}^{N-1}.
\ee 
Of these two sets of potential outcomes,
one is observed while the other is missing.As a result,
$H_0^A$ is not sharp,
and traditional randomization tests \citep{Fi41} are infeasible.

To test $H_0^A$, 
\citet{aronow2012general} introduced conditional randomization tests, 
which condition on the observed treatment assignments of a subset of focal units $\mathscr{U}$ and restrict $H_0^A$ to $\mathscr{U}$,
leading to a sharp null hypothesis.
To explain the idea of conditional randomization tests in one of the simplest possible settings,
sample one unit at random,
e.g.,
unit $1$.
If we condition on the observed treatment assignment $x_1^{obs}$ of unit $1$,
then
\bi 
\item the potential outcomes $y_1(1-x_1^{obs}; \bx_{N-1})$ for all $\bx_{N-1}\in\{0,1\}^{N-1}$ are eliminated,
\vspace{.1cm}
\item the potential outcomes $y_1(x_1^{obs}; \bx_{N-1}) = y_1^{obs}$ for all $\bx_{N-1}\in\{0,1\}^{N-1}$ are observed,
\ei 
where $y_1^{obs}$ is the observed outcome of unit $1$.
In the language of \citet{aronow2012general},
the set $\mU = \{1\}$ is a set of focal units.
A conditional randomization test can be conducted by randomizing the treatment assignments $X_2, \ldots, X_N$ of the non-focal units $\mathscr{U}^c = \{2, \ldots, N\}$ conditional on the observed treatment assignment $X_1 = x_1^{obs}$ of the focal unit $\mathscr{U} = \{1\}$.

In general,
the idea of conditional randomization tests applies to any non-empty set of focal units $\mU \subset \{1, \ldots, N\}$ sampled at random from $\{1, \ldots, N\}$.
Let $\bx^{obs}$ and $\by^{obs}$ be the observed treatment assignments and outcomes,
and $\mathscr{V}$ be the set of treatment assignments $\bx$ consistent with the observed treatment assignments $\bx^{obs}_{\mathscr{U}}$ of focal units $\mU$:
\be
\label{vref}
\mathscr{V} 
&\coloneqq& \{\bx\in\mX:\; \bx_{\mathscr{U}} = \bx^{obs}_{\mathscr{U}}\}.
\ee
Consider an $(\mathscr{U},\mathscr{V})$-restricted version of the null hypothesis $H_0^A$, 
called $H_0^{A,\mathscr{U},\mathscr{V}}$, 
which focuses on focal units $i \in \mathscr{U}$ with potential outcomes $y_i(\bx)$ ($\bx\in\mV$):
\beno
H_0^{A,\mathscr{U},\mathscr{V}}:\; y_i(\bx) = y_i(\bx')\, \text{ for all }\, i \in \mathscr{U},\, \bx,\, \bx'\in\mathscr{V}.
\ee
The null hypothesis $H_0^{A,\mathscr{U},\mathscr{V}}$ has two notable features.
First, 
testing $H_0^{A,\mathscr{U},\mathscr{V}}$ helps determine whether the potential outcomes $y_i(\bx)$ of focal units $i \in \mathscr{U}$ are affected by the treatment assignments $x_j$ of non-focal units $j\in \mathscr{U}^c$.
Second, 
$H_0^{A,\mathscr{U},\mathscr{V}}$ is sharp in the sense that,
for each focal unit $i\in\mU$, 
the potential outcomes $y_i(\bx)$ are identical for all $\bx\in\mV$ and are specified by $i$'s observed outcome $y_i^{obs}$,
as demonstrated by the example with $\mathscr{U} = \{1\}$ and $\mathscr{U}^c = \{2, \ldots, N\}$ described above.

Suppose that $T_{\mathscr{U},\mathscr{V}}(\bX, \by_{\mathscr{U}}(\bX)) \in \mR$ is a test statistic,
which is a function of the treatment assignments $\bX \in \mathscr{V}$ and corresponding potential outcomes $\by_{\mathscr{U}}(\bX)$ of focal units $\mathscr{U}$,
but does not depend on the potential outcomes $\by_{\mathscr{U}^c}(\bX)$ of non-focal units $\mathscr{U}^c$.
If $H_0^{A,\mathscr{U},\mathscr{V}}$ is true,
\beno
T_{\mathscr{U},\mathscr{V}}(\bx, \by_{\mathscr{U}}(\bx)) 
= T_{\mathscr{U},\mathscr{V}}(\bx, \by_{\mathscr{U}}^{obs})\ \text{for all } \bx\in\mathscr{V},
\ee
because the potential outcomes $\by_{\mathscr{U}}(\bx)$ of focal units $\mU$ are specified by their observed outcomes $\by_{\mathscr{U}}^{obs}$.
As a result,
the test statistic $T_{\mathscr{U},\mathscr{V}}(\bX, \by_{\mathscr{U}}(\bX))$ is identical to $T_{\mathscr{U},\mathscr{V}}(\bX, \by_{\mathscr{U}}^{obs})$ conditional on $\mathscr{U}$ and $\bX\in\mV$.
We can calculate $p$-values by randomizing the treatment assignments $\bX_{\mathscr{U}^c}$ of non-focal units $\mathscr{U}^c$ and evaluating how extreme the observed value $T_{\mU,\mV}(\bx^{obs},\by^{obs})$ of the test statistic $T_{\mU,\mV}(\bX,\by^{obs})$ is conditional on $\mU$ and $\bX \in \mathscr{V}$,
i.e.,
under the distribution of $\bX \mid \mathscr{U}, \bX \in \mathscr{V}$.
In experimental settings,
the distribution of $\bX \mid \mathscr{U}, \bX \in \mathscr{V}$ can be deduced from the distribution of $\bX$,
which is under the control of experimenters.
To present a simple example of how $p$-values can be calculated \citep{aronow2012general},
suppose that the probabilities $\mbP_{\mX\mid \mU,\mV}(\bX = \bx \mid \mU, \bX \in \mV)$ are identical for all $\bx \in \mathscr{V}$ and that large values of $T_{\mU,\mV}(\bX,\by^{obs})$ indicate evidence against $H_0^A$. 
Then the $p$-value is
\be
\label{p.value}
\dfrac{|\{\bx \in \mathscr{V}:\; T_{\mathscr{U},\mathscr{V}}(\bx, \by_\mathscr{U}^{obs})\, >\, T_{\mathscr{U},\mathscr{V}}(\bx^{obs}, \by_\mathscr{U}^{obs})\}|}{|\mathscr{V}|}.
\ee

In general,
two questions arise.
First, 
how should the set of focal units $\mU$ be sampled?
Second,
which test statistic $T_{\mathscr{U},\mathscr{V}}(\bX, \by_\mathscr{U}^{obs})$ should be chosen? 

The set of focal units $\mathscr{U}$ can affect the performance of conditional randomization tests of $H_0^A$, 
because such tests can only detect interference arising from the effect of treatment assignments $x_j$ of non-focal units $j \in \mathscr{U}^c$ on potential outcomes $y_i(\bx)$ of focal units $i \in \mathscr{U}$.
In some scenarios,
the design of the experiment may suggest a natural choice of focal units:
e.g.,
if the study involves households,
a natural choice of focal units would be to sample one member from each household \cite{athey2018exact}.
In other cases, 
focal units can be sampled at random from $\{1, \ldots, N\}$.

\citet{aronow2012general} suggests a test statistic capturing the dependence of $\bY_{\mathscr{U}}$ on $\bX_{\mathscr{U}^c}$,
e.g.,
the rank-correlation between $\bY_{\mathscr{U}}$ and $\bD_{\mathscr{U}}$,
where $\bD_{\mathscr{U}}$ are distances between units in $\mathscr{U}$ and the corresponding nearest treated units in $\mathscr{U}^c$.
An alternative is proposed by \citet{athey2018exact}:
\be
\label{athey}
T_{\mathscr{U},\mathscr{V}}(\bX, \by_\mathscr{U}^{obs})
\hspace{-.07cm}
\coloneqq 
\hspace{-.07cm}
|h(\bX, \by_\mathscr{U}^{obs})
\hspace{-.07cm}
- 
\hspace{-.07cm}
h(\bm{1} - \bX, \by_\mathscr{U}^{obs})|,
\ee
where
\beno
h(\bX,\, \by_\mathscr{U}^{obs})
\;\coloneq\; \dfrac{\sum_{i\in\mathscr{U}}\sum_{j\in\mathscr{U}^c} X_j\; y_i^{obs}\, z_{i,j}}{\sum_{i\in\mathscr{U}}\sum_{j\in\mathscr{U}^c} X_j\, z_{i,j}}.
\ee

\subsubsection{Detecting Specific Forms of Interference}
\label{sec:specific.interference}

Instead of testing whether there is any interference, investigators may be interested in testing for specific forms of interference: 
e.g.,
investigators may wish to test whether the potential outcomes of units at exposure levels $d_1$ and $d_2$ are identical:
\begin{gather*}
H_0^B:\; y_i(\bx) = y_i(\bx')\;\; \text{ for all }i,\, \bx,\, \bx'\text{ such that }f_i(\bx),f_i(\bx')\in\{d_1,d_2\}.
\end{gather*}
\citet{athey2018exact},
\citet{BaFeTo19},
\citet{PuBaFeTo21}, 
and others adapt the approach of \citet{aronow2012general} to test such hypotheses.
\citet{BaFeTo19} introduce a general procedure for detecting specific forms of interference based on conditional randomization tests and controlling the probability of type I error.
Consider any null hypothesis $H_0$ involving interference (e.g., 
$H_0^A$ or $H_0^B$).
Testing such hypotheses using conditional randomization tests is possible by choosing a conditioning event $\mathscr{E} \coloneqq (\mathscr{U}, \mathscr{V})$, 
where $\mathscr{U}$ is a non-empty subset of focal units sampled from $\{1, \ldots, N\}$ and $\mathscr{V}$ is a non-empty subset of treatment assignments that contains the observed treatment assignments $\bx^{obs}$.
The sampling design of $\mathscr{E}$ should ensure that the null hypothesis $H_0$ is sharp when restricted to $\mathscr{E}$, 
in the sense that
\be
\label{sharp}
y_i(\bx) = y_i(\bx')\, \text{ for all }\, i \in \mathscr{U},\ \bx,\, \bx'\in\mathscr{V}.
\ee
Equation \eqref{sharp} implies that,
for each focal unit $i\in\mU$,
the potential outcomes $y_i(\bx)$ are identical for all $\bx\in\mV$ and are specified by $i$'s observed outcome $y_i(\bx^{obs}) = y_i^{obs}$,
because $\bx^{obs}$ is contained in $\mV$.
An example of $(\mathscr{U}, \mathscr{V})$ for testing $H_0^A$ is described in Section \ref{sec:detect.interference}.

The following result is based on Theorem 1 of \citet{BaFeTo19}.

\begin{theorem}
\label{theorem.basse}
Let $H_0$ be a null hypothesis about interference (e.g., $H_0^A$ or $H_0^B$).
Sample $\mathscr{E} \coloneqq (\mathscr{U}, \mathscr{V})$ such that $H_0$ is sharp when restricted to $\mathscr{E}$ in the sense of Equation \eqref{sharp}.
Choose a test statistic $T_{\mathscr{U},\mathscr{V}}(\bX, \by_\mathscr{U}^{obs})$ that is a function of treatment assignments $\bX$ and observed outcomes $\by_\mathscr{U}^{obs}$ of focal units $\mathscr{U}$,
and calculate the $p$-value
\beno
p(\bx^{obs}) 
\coloneqq 
\mbP_{\mX\mid\mathscr{E}}(T_{\mathscr{U},\mathscr{V}}(\bX, \by_\mathscr{U}^{obs}) 
> 
T_{\mathscr{U},\mathscr{V}}(\bx^{obs}, \by_\mathscr{U}^{obs}) \mid \mathscr{E})
\ee
under the conditional distribution of $\bX \mid \mathscr{E}$.
If $H_0$ is true,
\beno 
\mbP_{\mX\mid\mathscr{E}}(p(\bX^{obs}) \leq \alpha \mid \mathscr{E}) \leq \alpha\;
\mbox{ for all } \alpha \in [0, 1].
\ee
\end{theorem}

\vspace{-1cm}

\noindent
Theorem \ref{theorem.basse} does not make assumptions about the test statistic.
Having said that,
the $p$-value stated in Theorem \ref{theorem.basse} is meaningful when large values of the test statistic indicate evidence against $H_0$. 
An example of a test statistic for testing $H_0^A$ is shown in Equation \eqref{athey},
and an example of a $p$-value can be found in Equation \eqref{p.value}.

\vspace{.1cm}

\noindent{\em Implications.}
Suppose that we observe treatment assignments $\bx^{obs}$ and outcomes $\by_\mathscr{U}^{obs}$ of focal units $\mathscr{U}$,
we pick $\alpha \in (0, 1)$,
and we make the following decision:
\bi 
\item If $p(\bx^{obs}) \leq \alpha$, 
we reject $H_0$.
\vspace{0.1cm}
\item If $p(\bx^{obs}) > \alpha$, 
we do not reject $H_0$.
\ei 
If $H_0$ is true and $p(\bx^{obs}) \leq \alpha$,
we reject $H_0$ and commit an error of type I.
Theorem \ref{theorem.basse} guarantees that the probability of committing an error of type I is at most $\alpha$:
\beno 
\mbP_{\mX\mid\mathscr{E}}(p(\bX^{obs}) \leq \alpha \mid \mathscr{E}) 
&\leq& \alpha.
\ee
The probability of type I error is based on the probability law $\mbP_{\mX\mid\mathscr{E}}$,
because the $p$-value is based on the distribution of treatment assignments $\bX$ conditional on the event $\mathscr{E}$ of sampling $\mathscr{U}$ and observing $\bX \in \mathscr{V}$.
Theorem \ref{theorem.basse} controls the probability of type I error regardless of which test statistic and $\mathscr{E}$ are chosen.
That said,
the choice of test statistic and $\mathscr{E}$ can affect the probability of type II error and hence the power of the test.

In the tests of \citet{aronow2012general} and \citet{athey2018exact},
the choice of focal units $\mathscr{U}$ is independent of the observed treatment assignments $\bx^{obs}$.
\citet{BaFeTo19} demonstrate that sampling $\mathscr{U}$ dependent on $\bx^{obs}$ can increase the power of tests.
For example, 
if investigators wish to test $H_0^B$,
it is advisable to sample focal units from the set of units whose outcomes contain information about $H_0^B$,
i.e.,
units $i$ with observed exposure levels $f_i(\bx^{obs}) \in \{d_1, d_2\}$. 
Following \citet{BaFeTo19}, 
\citet{PuBaFeTo21} propose a general method to constructing $\mathscr{E} \coloneqq (\mathscr{U}, \mathscr{V})$ using a graph-theoretic approach, 
and suggest guidelines on how to construct tests with high power.
\citet{online.experiment} develop conditional randomization tests for detecting interference in online controlled experiments with increasing treatment allocation.

Last,
but not least,
conditional randomization tests are based on fixed potential outcomes $\by(\cdot)$.
If potential outcomes $\bY(\cdot)$ are random, 
these tests can be interpreted as being applied to a realization $\by(\cdot)$ of $\bY(\cdot)$.
In other words:
if potential outcomes $\by(\cdot)$ are generated from a probability law governing $\bY(\cdot)$,
the described tests can be applied conditional on the set of realized potential outcomes $\bY(\cdot)=\by(\cdot)$;
note that all potential outcomes with the exception of $\by(\bx^{obs}) = \by^{obs}$ are unobserved,
but the observed outcomes $\by^{obs}$ are sufficient for calculating test statistics $T_{\mathscr{U},\mathscr{V}}(\bX, \by_{\mathscr{U}}^{obs})$ and $p$-values under the distribution of $\bX \mid \mathscr{U}, \bX \in \mathscr{V}$.
The tests remain valid under ignorability or conditional ignorability, 
which ensure that the probability law of treatment assignments $\bX$ is invariant conditional on $\bY(\cdot)=\by(\cdot)$.

\subsection{Tests for Average Causal Effects}
\label{sec:statistical.tests.causal.effects}

The conditional randomization tests discussed in Section \ref{sec:conditional.randomization.test.interference} help detect interference at the individual level:
e.g.,
$H_0^A$ claims that no unit is affected by any other,
whereas $H_1^A$ states that at least one unit is affected by others.
Instead of testing whether there is no interference at all,
investigators may be interested in testing whether the average causal effect across units vanishes:
e.g.,
if $H_0^A$ is true,
the indirect causal effect $\tau^I_{\mX}$ defined in Section \ref{sec:causal.effects.fixed} vanishes,
but the converse is not true. 
If $H_0^C: \tau^I_{\mX}=0$ is true and $H_0^C$ was tested by using the tests described in Section \ref{sec:conditional.randomization.test.interference},
investigators would commit many errors of type I,
because the tests described in Section \ref{sec:conditional.randomization.test.interference} would often reject $H_0^C$ when there is interference but the average causal effect vanishes.
More appropriate tests for average causal effects 
can be constructed by leveraging finite- or large-sample results for estimators of causal effects:
e.g.,
\citet{choi2017estimation} provides finite- or large-sample confidence intervals that can be used to construct tests for average causal effects,
while \citet{aronow2017estimating} and \citet{LiWa22} report asymptotic normality results.

\vspace{.1cm}

\subsection{Simulation Results}
\label{sec:simulation}

We shed light on the performance of the interference tests described in Section \ref{sec:conditional.randomization.test.interference} using simulations. 

\vspace{.1cm}

{\em Initialization.}
To initialize the simulation study,
we need a set of potential outcomes $\{\by(\bx): \bx \in \mX\}$ and an interference graph $\bz$.
Given an interference graph $\bz$,
there are at least two approaches to initializing the set of potential outcomes $\{\by(\bx): \bx \in \mX\}$:
\begin{enumerate}
\item For each $\bx \in \mX$,
specify potential outcomes $\by(\bx)$ by specifying a deterministic function $g(\bx; \bz)$ of $\bx$ given $\bz$ and evaluating $g(\bx; \bz)$.
\vspace{.05cm}
\item For each $\bx \in \mX$, 
generate potential outcomes by specifying a probability law of $\bY(\bx)$ with expectation $\mbE_{\mY(\bx)\mid\mZ}[\bY(\bx) \mid \bZ = \bz] = g(\bx; \bz)$ and generating a realization $\by(\bx)$ of $\bY(\bx)$.
\end{enumerate}
\vspace{.05cm}
Both approaches require assumptions about $g(\bx; \bz)$.
Since the test procedure described in Section \ref{sec:conditional.randomization.test.interference} does not make assumptions about the process generating $\{\by(\bx): \bx \in \mX\}$,
either the first or second approach can be used to initialize $\{\by(\bx): \bx \in \mX\}$. 
We use the second approach because it allows random deviations from $g(\bx; \bz)$.

To generate a set of potential outcomes $\{\by(\bx): \bx \in \mX\}$,
we generate realizations $\by(\bx)$ of $\bY(\bx)$:
\beno
\bY(\bx) 
= \beta\, \bx + \gamma\, \bC_{N,1}(\bz)\, \bz\, \bx + \delta\, \bC_{N,2}(\bz)\, \bz\, \bY(\bx) + \bm{\epsilon},
\ee
where $\beta \in \mR$,
$\gamma \in \mR$,
and $\delta \in (-1, +1)$ can be interpreted as the effects of treatment,
spillover,
and contagion,
respectively;
$\bC_{N,1}(\bz)$ and $\bC_{N,2}(\bz)$ are diagonal matrices with diagonal elements $c_{N,i,1}(\bz) \coloneqq 1/\sum_{j=1}^N\, z_{i,j}$ and $c_{N,i,2}(\bz) \coloneqq 1/\mnorm{\bz}_2$,
respectively;
and $\bm{\epsilon}$ is multivariate Gaussian with mean $\bm{0}_N$ and covariance matrix $(\bm{I}_N - \delta\,\bC_{N,2}(\bz)\,\bz)$,
which is positive-definite.

The interference graph $\bz$ is generated by a rank-one random graph model \citep{latent.space.models.theory}:
\beno 
Z_{i,j} 
&\ind& \text{Bernoulli}(P_{i,j}),\;
& 
P_{i,j} 
&\coloneqq& \alpha_i\, \alpha_j,\;
& \alpha_i\, \iid\, \mbox{Beta}(1,\, 3).
\ee

\vspace{.1cm}

{\em Simulation procedure.}
Given the generated set of potential outcomes $\{\by(\bx): \bx \in \mX\}$ and interference graph $\bz$,
we simulate $R =$ 1,000 replications of $(\bx^{(r)},\by^{(r)})$ $(r=1,\ldots, R)$.
For the $r$th replication, 
we first generate treatment assignments $\bx^{(r)}$ independently from
\beno
\mbP_{\mX}(\bX = \bx)
&=& \dfrac{1}{\binom{N}{M}}\ \text{for all }\bx\text{ such that}\ \dsum_{i=1}^{N} x_i = M,
\ee
where $M \coloneqq 2\, N / 5$. 
Based on the generated treatment assignments $\bx^{(r)}$, 
we observe outcomes $\by^{(r)}=\by(\bx^{(r)})$.
We select $30\%$ of the observations at random and use them as focal units $\mathscr{U}$.
Given $\mathscr{U}$,
we construct $\mathscr{V}$ defined in Equation \eqref{vref}.
We then test $H_0^{A}$ using the test statistic $T_{\mathscr{U},\mathscr{V}}(\bX, \by^{(r)}_\mathscr{U})$ in Equation \eqref{athey} and compute the $p$-value in Equation \eqref{p.value} under the distribution of $\bX\mid \mU,\bX\in\mV$.

\vspace{.1cm}

{\em Simulation scenarios.}
We consider the scenarios
\begin{itemize}
\item[(a)] $\beta \neq 0,\,\gamma \neq 0,\,\delta=0$ (spillover, no contagion);
\vspace{.1cm}
\item[(b)] $\beta \neq 0,\,\gamma\neq 0,\,\delta\neq 0$ (spillover and contagion);
\vspace{.1cm}
\item[(c)] $\beta \neq 0,\,\gamma=0,\,\delta\neq 0$ (contagion, no spillover).
\end{itemize}
We use $\beta = 5$,
$\gamma \in \{0, 5\}$,
and $\delta \in \{0,\, 0.25,\, 0.5,\, 0.75\}$. 
For each combination of $(\beta,\gamma,\delta)$ and $N \in \{100, 500\}$,
we generate data using the above simulation procedure.

\vspace{.1cm}

\begin{figure}
\centering
\includegraphics[width=0.6\linewidth]{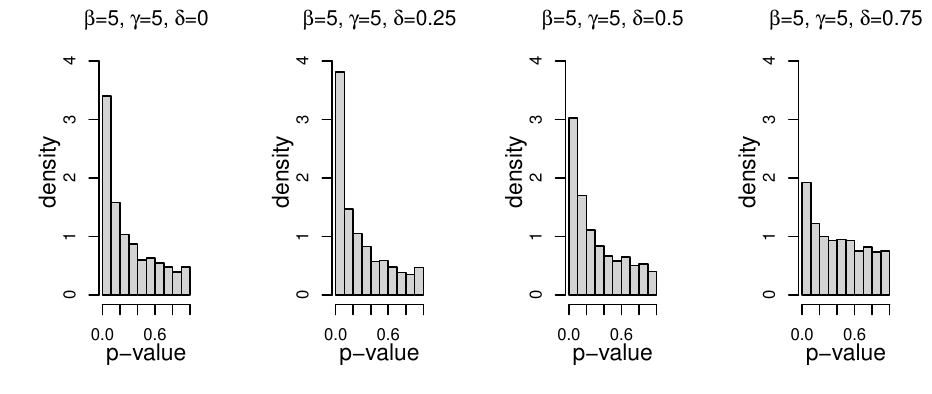}
\includegraphics[width=0.6\linewidth]{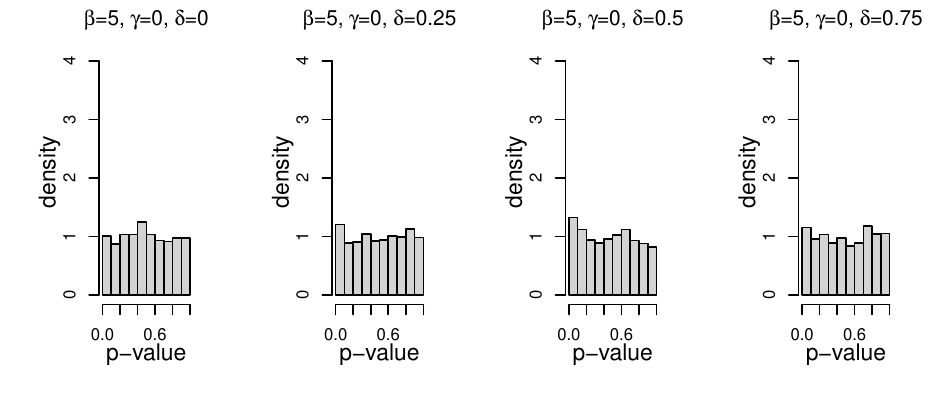}
\caption{Simulation results based on $N=100$ units: 
$p$-values based on the test statistic in Equation \eqref{athey} with treatment effect $\beta$, spillover effect $\gamma$, and contagion effect $\delta$.
} 
\label{fig:test}
\end{figure}

\begin{figure}
\centering
\includegraphics[width=0.6\linewidth]{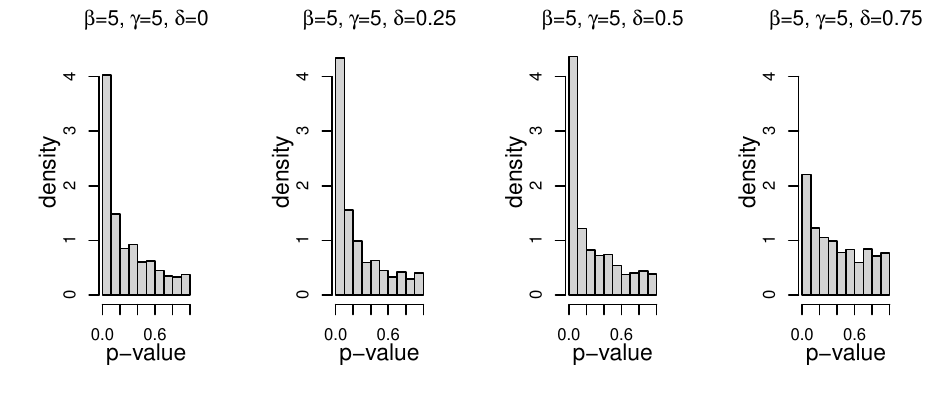}
\includegraphics[width=0.6\linewidth]{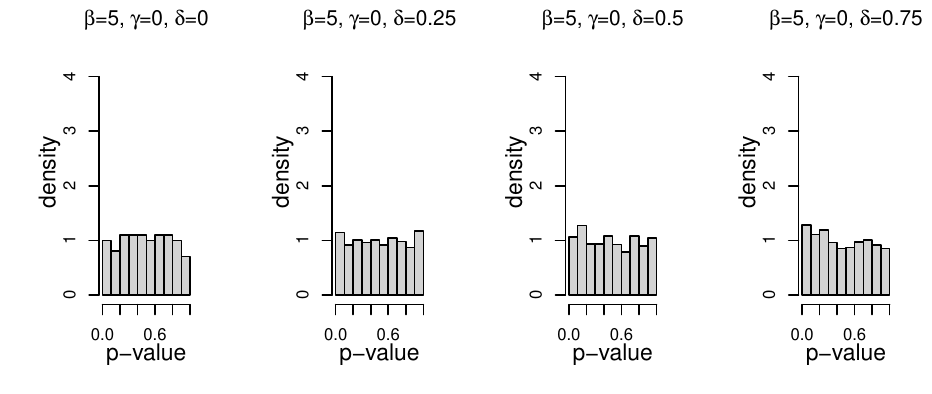}
\caption{Simulation results based on $N = 500$ units: 
$p$-values based on the test statistic in Equation \eqref{athey} with treatment effect $\beta$, spillover effect $\gamma$, and contagion effect $\delta$. 
} 
\label{fig:test2}
\end{figure}

{\em Simulation results.}
The resulting $p$-values are shown in Figures \ref{fig:test} and \ref{fig:test2}.
In the first and second scenario with spillover ($\gamma \neq 0$),
the test is effective at detecting interference. 
In the third scenario with contagion ($\delta \neq 0$) but without spillover ($\gamma = 0$),
the test is less effective at detecting interference;
increasing $N$ from 100 to 500 does not improve test performance.

\vspace{.1cm}

{\em Discussion and open questions.}
Spillover implies interference,
as does contagion (Section \ref{sec:capturing.contagion}):
spillover implies direct interference,
whereas contagion implies indirect interference.
The simulation results suggest that the test statistic in Equation \eqref{athey} is effective at detecting direct interference,
but may be less effective at detecting indirect interference.
To detect indirect interference,
other test statistics may be preferable.
It is worth noting that the interference testing framework established in \citet{aronow2012general},
\citet{athey2018exact},
\citet{BaFeTo19},
\citet{PuBaFeTo21},
and others allows researchers to choose from a wide range of test statistics.
Which test statistics are well-suited to detecting indirect interference is an open question.
The test statistic in Equation \eqref{athey} may not be well-suited,
because it is based on sums of the form $\sum_{i\in\mathscr{U}} \sum_{j\in\mathscr{U}^c} X_j\,\, y_i^{obs}\, z_{i,j}$,
i.e.,
it captures interference via direct neighbors (neighborhood exposure).
Test statistics that are not based on neighborhood exposure assumptions are promising alternatives.

\section{Estimating Causal Effects}
\label{sec:estimation}

A central objective of causal inference is to find a function of observed outcomes $\bY$ for estimating causal effects based on potential outcomes $\bY(\cdot)$.
We review 
\bi 
\item design-based estimators,
including unbiasedness of design-based estimators and variance reduction methods (Section \ref{sec:design.based.estimators});
\vspace{.05cm}
\item model-based estimators (Section \ref{sec:model.based.estimators});
\vspace{.05cm}
\item unknown interference graphs (Sections \ref{sec:unknown.interference});
\vspace{.05cm}
\item exposure misspecification (Section \ref{sec:exposure.misspecification}).
\ei 
{\blues
Throughout Section \ref{sec:estimation},
we assume that causal effects are identified.
Identification is discussed in Section \ref{sec:identification}.

\subsection{Design-Based Estimators}
\label{sec:design.based.estimators}

Motivated by the observation that many researchers wish to avoid assumptions about potential outcomes and interference graphs,
we call an estimator design-based as long as
it avoids assumptions about the process generating potential outcomes and interference graphs.
A design-based estimator can be based on assumptions about the design of the experiment or observational study---including treatment assignments $\bX$---as long as it does not rely on assumptions about potential outcomes and interference graphs.
Design-based estimators can be used when potential outcomes and interference graphs are considered fixed or random.
We first discuss various design-based estimators,
and then review
\bi 
\item unbiasedness and asymptotic unbiasedness of design-based estimators (Section \ref{sec:unbiasedness});
\vspace{.05cm}
\item variance reduction methods for design-based estimators (Section \ref{sec:variance.reduction}).
\ei

A well-known class of design-based estimators consists of Horvitz-Thompson estimators,
sometimes called Inverse Propensity Weighted estimators.
Without interference, 
the Horvitz-Thompson estimator of the average treatment effect $\tau$ defined in Section \ref{sec:target} is
\beno
\widehat\tau_{HT}
&\coloneqq& \dfrac{1}{N} \dsum_{i=1}^N \left(\dfrac{Y_i\; \mbI(X_i = 1)}{\mbP_\mX(X_i = 1)} - \dfrac{Y_i\; \mbI(X_i = 0)}{\mbP_\mX(X_i = 0)}\right).
\ee
In the absence of interference,
the Horvitz-Thompson estimator $\widehat\tau_{HT}$ is an unbiased estimator of the average treatment effect $\tau$.
With interference, 
the conditional Horvitz-Thompson estimator
\beno
\widehat\tau_{HT}^D
\coloneqq 
\dfrac{1}{N}
\dsum_{i=1}^N 
\left(\dfrac{Y_i\ \mbI(X_i = 1)}{\mbP_{\mX\mid\mZ}(X_i = 1 \mid \bZ)} 
- 
\dfrac{Y_i\, \mbI(X_i = 0)}{\mbP_{\mX\mid\mZ}(X_i = 0 \mid \bZ)}\right)
\ee
is an unbiased estimator of the direct causal effects $\tau^D_{\mX}$ and $\tau^D_{\mX,\mY(\cdot),\mZ}$ defined in Sections \ref{sec:causal.effects.fixed} and \ref{sec:causal.effects.random}.
We establish unbiasedness of $\widehat\tau_{HT}^D$ in Section \ref{sec:unbiasedness}.

An alternative to the Horvitz-Thompson estimator is the H\'ajek estimator,
which replaces the treatment assignment probabilities $\mbP_{\mX\mid\mZ}(X_i = 1\mid \bZ=\bz)$ by moment estimators and can be useful when the treatment assignment probabilities are unknown.
The H\'ajek estimator of $\tau^D_\mX$ is
\beno
\widehat\tau_{HAJ}^D
&\coloneqq& 
\dfrac{\sum_{i=1}^N Y_i\ \mbI(X_i = 1)}{\sum_{i=1}^N \mbI(X_i = 1)}
-
\dfrac{\sum_{i=1}^N Y_i\ \mbI(X_i = 0)}{\sum_{i=1}^N \mbI(X_i = 0)}.
\ee
\citet{savje2021average} and \citet{LiWa22} compare Horvitz-Thompson and H\'ajek estimators of $\tau^D_\mX$.
\vspace{-.4cm}

Causal effects based on exposure mappings \citep{aronow2017estimating} can be estimated by Horvitz-Thompson estimators as well:
e.g.,
the average potential outcome at exposure level $d$,
\beno
\label{avg.exposure.effect}
\mu^E(d) 
&\coloneqq& \dfrac{1}{N}\dsum_{i=1}^N Y_i^E(d),
\ee
can be estimated by the Horvitz-Thompson estimator
\beno
\widehat\mu_{HT}^E(d)
&\coloneqq& \dfrac{1}{N} \dsum_{i=1}^N \dfrac{Y_i\, \mbI(f_i(\bX) = d)}{\mbP_{\mcF_i\mid \mZ}(f_i(\bX) = d \mid \bZ)}.
\ee
In experimental settings,
the exposure probabilities
$\mbP_{\mcF_i\mid \mZ}(f_i(\bX) = d \mid \bZ = \bz)$ are known, and the Horvitz-Thompson estimator is an unbiased estimator of $\mu^E(d)$ under ignorability.
In observational settings with observed confounders $\bC_i$, 
the exposure probabilities $\mbP_{\mcF_i\mid\mZ}(f_i(\bX) = d \mid \bZ = \bz)$ can be replaced by propensity scores \cite{forastiere2021identification}:
\beno
\psi(d;\, \bc, \bz)
&\coloneqq& \mbP_{\mcF_i\mid \mC_i,\mZ}(f_i(\bX) = d \mid (\bm{C}_i,\bZ)=(\bc,\bz)).
\ee
If the number of exposure levels is finite,
one can estimate propensity scores using multinomial regression.
Then $\mu^E(d)$ can be estimated by
\beno
\widehat\mu_{HT}^E(d) 
&\coloneqq& \dfrac{1}{N}\dsum_{i=1}^N \dfrac{Y_i\; \mbI(f_i(\bX) = d)}{\widehat\psi(d; \bm{C}_i, \bZ)},
\ee
where $\widehat\psi(d; \bC_i, \bZ)$ is an estimator of $\psi(d; \bC_i, \bZ)$.
Note that propensity score estimators can be viewed as design-based:
while propensity scores rely on the probability law of treatment assignments $\bX$,
they do not make assumptions about the process generating potential outcomes and interference graphs $(\bY(\cdot), \bZ)$.
An alternative to the Horvitz-Thompson estimator $\widehat\mu_{HT}^E(d)$ is the H\'ajek estimator $\widehat\mu_{HAJ}^E(d)$:
\beno
\widehat\mu_{HAJ}^E(d)
&\coloneqq& \dfrac{\sum_{i=1}^N Y_i\; \mbI(f_i(\bX) = d)}{\sum_{i=1}^N \mbI(f_i(\bX) = d)}.
\ee
Horvitz-Thompson and H\'ajek estimators of $\mu^E(d)$ require a large enough number of observations at exposure level $d$ to be consistent \cite{forastiere2021identification,leung2020treatment,gao2025regression}.
If the number of exposure levels is large,
it may not be possible to have enough observations at all exposure levels.
To address these sample size issues,
\citet{forastiere2021identification} suggested sub-classification techniques.

\subsubsection{Unbiasedness}
\label{sec:unbiasedness}

Many design-based estimators are unbiased or asymptotically unbiased.
As a case in point,
we consider the Horvitz-Thompson estimator $\widehat\tau_{HT}^D$ defined in Section \ref{sec:design.based.estimators}.
We show that under Bernoulli randomization designs $\widehat\tau_{HT}^D$ is an unbiased estimator of the direct causal effect $\tau^D_{\mX}$ based on fixed potential outcomes and interference graphs $(\by(\cdot),\bz)$,
as well as the direct causal effect $\tau^D_{\mX,\mY(\cdot),\mZ}$ based on random potential outcomes and interference graphs $(\bY(\cdot),\bZ)$.
To establish unbiasedness of $\widehat\tau_{HT}^D$,
recall that causal effects based on $(\by(\cdot),\bz)$ can be viewed as conditional causal effects given $(\bY(\cdot),\bZ) = (\by(\cdot),\bz)$ (Section \ref{sec:from.fixed.to.random}).
Thus,
for any unit $i$ and treatment assignment $x_i \in \{0, 1\}$,
\beno
\mbE_{\mX\mid \mY(\cdot),\mZ}\left[\dfrac{Y_i\, \mbI(X_i = x_i)}{\mbP_{\mX\mid\mZ}(X_i = x_i \mid \bZ = \bz)} \;\bigg|\; (\bY(\cdot), \bZ) = (\by(\cdot), \bz)\right]\s
\\
= \dfrac{\mbE_{\mX\mid\mY(\cdot),\mZ}[y_i(x_i;\bX_{-i}) \mbI(X_i=x_i) \;\big|\; (\bY(\cdot), \bZ) =  (\by(\cdot), \bz)]}{\mbP_{\mX\mid\mZ}(X_i = x_i \mid \bZ = \bz)}
\ee
by the consistency assumption in Equation \eqref{consistency.assumption},
and
\beno
&=& \mbE_{\mX\mid\mZ}[y_i(x_i; \bX_{-i}) \mid \bZ = \bz]
\ee
by ignorability and the independence of $X_i$ and $\bX_{-i}$ 
under Bernoulli randomization designs.
Thus,
$\widehat\tau_{HT}^D$ is an unbiased estimator of the causal effect $\tau^D_\mX$ based on fixed potential outcomes and interference graphs:
\beno
\mbE_{\mX\mid \mY(\cdot),\mZ}\, \widehat\tau_{HT}^D 
&=& \tau^D_\mX.
\ee
In addition,
$\widehat\tau_{HT}^D$ is an unbiased estimator of the causal effect $\tau^D_{\mX,\mY(\cdot),\mZ}$ based on random potential outcomes and interference graphs:
\beno
\mbE_{\mX,\mY(\cdot),\mZ}\, \widehat\tau_{HT}^D 
&=& \mbE_{\mY(\cdot),\mZ}\, [\, \mbE_{\mX\mid\mY(\cdot),\mZ}\, \widehat\tau_{HT}^D]&=& \mbE_{\mY(\cdot),\mZ}\, \tau^D_\mX &=& \tau^D_{\mX,\mY(\cdot),\mZ}.
\ee

\subsubsection{Variance Reduction}
\label{sec:variance.reduction}

Design-based estimators can have a large variance under Bernoulli or complete randomization designs.
As a result,
the development of variance reduction methods for design-based estimators is an important area of research.
One line of research on variance reduction builds on cluster randomization designs,
starting with \citet{ugander2013graph},
\citet{eckles2017design},
and \citet{leung2022rate,leung2026cluster}.
In cluster randomization designs, units are assigned to clusters and treatments are allocated to entire clusters of units rather than individual units.
Suppose that each unit belongs to one of $K \ll N$ disjoint clusters $\mathscr{G}_1, \ldots, \mathscr{G}_K$.
Then,
for each cluster,
all units in the cluster are either assigned to the treatment group or the control group.
To construct clusters, 
investigators can leverage auxiliary information (i.e., covariates) or estimate clusters based on the interference graph. 
To demonstrate how cluster-level treatment assignments can reduce the variance of design-based estimators,
\citet{ugander2013graph} use the causal effect $\tau(\bm{1}_N, \bm{0}_N)$ defined in Section \ref{sec:causal.effects.fixed}.
Assume that the following neighborhood exposure assumption is satisfied:
\beno
\text{If}\ (x_i, \bx_{\mcN_i}) = (x'_i, \bx'_{\mcN_i}),\ \text{then }y_i(\bx) = y_i(\bx'),
\ee
where $\mcN_i\coloneqq\{j:z_{i,j}=1\}$ is the neighborhood of unit $i$.
A Horvitz-Thompson estimator of $\tau(\bm{1}_N, \bm{0}_N)$ is
\beno
\widehat\tau_{HT}(\bm{1}_N, \bm{0}_N) 
&\coloneqq& \dfrac{1}{N} \dsum_{i=1}^N \dfrac{Y_i\, \mbI((X_i, \bX_{\mcN_i}) = (1, \bm{1}_{|\mcN_i|}))}{\mbP_{\mX}((X_i, \bX_{\mcN_i}) = (1, \bm{1}_{|\mcN_i|}))}\s
\\
&-& \dfrac{1}{N} \dsum_{i=1}^N \dfrac{Y_i\, \mbI((X_i, \bX_{\mcN_i}) = (0, \bm{0}_{|\mcN_i|}))}{\mbP_{\mX}((X_i, \bX_{\mcN_i}) = (0, \bm{0}_{|\mcN_i|}))}.
\ee
If $X_i\, \iid\, \mbox{Bernoulli}(\pi)$ and $\pi \in (0, 1)$ is a constant independent of $N$,
the exposure probabilities
\beno 
\mbP_{\mX}((X_i, \bX_{\mcN_i}) = (1, \bm{1}_{|\mcN_i|})) 
&\;=\;& \pi^{|\mcN_i|+1}\s 
\\
\mbP_{\mX}((X_i,\bX_{\mcN_i}) = (0, \bm{0}_{|\mcN_i|}))
&\;=\;& (1-\pi)^{|\mcN_i|+1}
\ee
decay exponentially fast to $0$ if the neighborhood size $|\mcN_i|$ increases without bound as $N \to \infty$.
In practice,
given a data set with $N$ units,
neither $|\mcN_i|$ nor $N$ increase without bound,
but the above observation has real-world implications:
if some units have large neighborhoods,
the exposure probabilities in the denominator of $\widehat\tau_{HT}(\bm{1}_N, \bm{0}_N)$ can be small and the variance of $\widehat\tau_{HT}(\bm{1}_N, \bm{0}_N)$ can be large.
Cluster-level treatment assignments help reduce the variance of $\widehat\tau_{HT}(\bm{1}_N, \bm{0}_N)$:
e.g.,
if unit $i$ and its neighbors belong to $S_i \leq K$ clusters,
the exposure probabilities are
\beno 
\mbP_{\mX}((X_i, \bX_{\mcN_i}) = (1, \bm{1}_{|\mcN_i|})) 
&\;=\;& \pi^{S_i}\s 
\\
\mbP_{\mX}((X_i,\bX_{\mcN_i}) = (0, \bm{0}_{|\mcN_i|}))
&\;=\;& (1-\pi)^{S_i}.
\ee
If $\pi$ and $K$ are constants independent of $N$,
these exposure probabilities cannot be arbitrarily close to $0$,
which demonstrates that cluster-level treatment assignments help reduce the variance of Horvitz-Thompson estimators of $\tau(\bm{1}_N, \bm{0}_N)$.

To reduce the variance of Horvitz-Thompson estimators for the average exposure effect $\mu^E(d)$, 
\citet{aronow2017estimating} recommend using covariates.
If investigators have prior knowledge about covariates that may affect potential outcomes, 
the Horvitz-Thompson estimator can be adjusted using these covariates without compromising unbiasedness \citep{Raj65}.
\citet{LiWa22} develop a related idea to reduce the variance of a Horvitz-Thompson estimator of the indirect effect $\tau^I_{\mX}$, 
where unobserved attributes---such as community memberships of units in the interference graph---are estimated from an observed interference graph.
Other examples include a randomization design based on quasi-coloring by \citet{jagadeesan2020designs} and a two-wave experimental design by \citet{viviano2020experimental}.
\citet{SuAi17} focus on linear design-based estimators and propose an optimality criterion to achieve minimum variance.
\citet{kandiros2026conflict} introduce a unifying approach to constructing experimental designs for estimating a pre-specified causal effect 
based on the notion of conflict graphs and provide bounds on the variance of Horvitz-Thompson estimators.

\subsection{Model-Based Estimators}
\label{sec:model.based.estimators}

More often than not,
design-based estimators have been used in scenarios in which potential outcomes are fixed, 
with some notable exceptions \citep[e.g.,][]{leung2020treatment,LiWa22}.
If potential outcomes are random and generated by a model,
one can use model-based estimators.
Examples include maximum likelihood and Bayesian estimators \citep{ToKa13,FoMeWuAi22},
maximum pseudo-likelihood estimators \citep{TTFuSh21,BhSe24},
and targeted maximum likelihood estimators \citep{ogburn2024causal}.

A popular example is doubly robust estimators based on exposure mappings \citep{liu2019doubly,viviano2024policy,khatami2025gml,bong2025heterogeneous}.
Doubly robust estimators combine design-based estimators with model-based estimators that make assumptions about the process generating potential outcomes,
and can hence be viewed as model-based.
Let $m(d; \bm{c}_i,\bz)$ be the conditional expectation of the outcome $Y_i$ of unit $i$ given exposure level $f_i(\bX) = d$,
confounders $\bm{C}_i = \bm{c}_i$,
and interference graph $\bZ = \bz$:
\beno
m(d; \bc_i, \bz) 
&\coloneqq& \mbE_{\mY\mid\mathscr{F}_i,\mC_i,\mZ}[Y_i \mid (f_i(\bX), \bm{C}_i, \bZ) = (d, \bc_i, \bz)].
\ee
A doubly robust estimator of $\mu^E(d)$ is
\beno
\widehat\mu_{DR}^E(d)
&\coloneqq& \dfrac{1}{N} \dsum_{i=1}^N \dfrac{(Y_i - \widehat{m}(d; \bm{C}_i, \bZ))\, \mbI(f_i(\bX) = d)}{\widehat\psi(d; \bm{C}_i, \bZ)} &+& \dfrac{1}{N} \dsum_{i=1}^N \widehat{m}(d; \bm{C}_i, \bZ),
\ee
where $\widehat\psi(d; \bm{C}_i, \bZ)$ is an estimator of the propensity score $\psi(d; \bm{C}_i, \bZ)$ and $\widehat{m}(d; \bC_i, \bZ)$ is an estimator of the conditional expectation $m(d; \bC_i, \bZ)$ obtained by regressing the observed outcomes $Y_i$ on $(f_i(\bX), \bm{C}_i, \bZ) = (d, \bc_i, \bz)$ ($i = 1, \ldots, N$).
The estimator $\widehat\mu_{DR}^E(d)$ combines a design-based estimator $\widehat\psi(d; \bm{C}_i, \bZ)$ with a model-based estimator $\widehat{m}(d; \bC_i, \bZ)$,
which makes assumptions about the process generating potential outcomes $\bY(\cdot)$ to estimate $\widehat{m}(d; \bC_i, \bZ)$.
Its double robustness stems from the fact that if either the design-based estimator $\widehat\psi(d; \bC_i, \bZ)$ or the model-based estimator $\widehat{m}(d; \bC_i, \bZ)$ is consistent, 
then so is $\widehat\mu_{DR}^E(d)$ \citep{viviano2024policy}.

}

\subsection{Unknown Interference}
\label{sec:unknown.interference}

It is possible that neither interference graphs nor exposure mappings can be specified.
In such scenarios,
\citet{choi2017estimation} proposes confidence intervals to upper bound
\beno 
\mu(\bm{0}_N)
&\coloneqq& \dfrac1N \dsum_{i=1}^N y_i(\bm{0}_N),
\ee
which lower bounds the attributable average treatment effect \cite{rosenbaum2001effects}:
\beno
\dfrac{1}{N} \dsum_{i=1}^N Y_i - \mu(\bm{0}_N),
\ee
assuming that potential outcomes $y_i(\bx)$ are fixed.
The proposed method relies on the monotone treatment effect assumption $y_i(\bx) \geq y_i(\bm{0}_N)$ for all $\bx \in \mX$ and uses optimization to compute estimates. 
The resulting estimates can be conservative.

\citet[][p.~2]{yu2022estimating} address the problem of unobserved network effects under the spillover model
\beno
y_i(\bx)
&\coloneqq& \alpha_i + \beta_i\, x_i + \dsum_{j=1,\, j \neq i}^N \gamma_{i,j}\, x_j,
\ee
where $\alpha_i \in \mR$,
$\beta_i \in \mR$,
and $\gamma_{i,j} \in \mR$.
\citet{yu2022estimating} study linear estimators for the causal effect $\tau(\bm{1}_N, \bm{0}_N)$ and propose linear unbiased estimators,
provided that information about $\mu(\bm{0}_N)$ is available.

\citet{savje2021average} established rates of convergence for Horvitz-Thompson and H\'ajek estimators for the direct effect $\tau^D_{\mX}$ defined in Section \ref{sec:causal.effects.fixed} under restrictions on interference, 
without requiring knowledge of a specified interference graph.
\citet{egami2021spillover} study causal inference in settings where interference can occur through multiple networks, 
some of which are unobserved.
While estimating direct causal effects is possible,
estimating indirect causal effects without knowledge of the interference graph is more challenging.

\subsection{Exposure Misspecification}
\label{sec:exposure.misspecification}

Exposure mappings help construct estimators and draw causal conclusions, 
but an important question is how sensitive causal conclusions are to misspecifications of exposure mappings.
{\blues
\citet{leung2022causal} and \citet{savje2024causal} study the properties of the Horvitz-Thompson estimator of the causal effect $\tau^E(d_1, d_2) \coloneqq \mu^E(d_1) - \mu^E(d_2)$ under exposure misspecification when potential outcomes $\by(\cdot)$ are fixed.

Under misspecified exposures,
$\mu^E(d)$ is not well-defined, 
because the potential outcomes $y_i(\cdot)$ can no longer be represented as $y_i^E(d)$.}
Instead, 
one can consider the following modified version:
\beno
\widetilde\tau^E(d_1, d_2)
&\coloneqq& \widetilde\mu^E(d_1) - \widetilde\mu^E(d_2),
\ee
where
\beno
\widetilde\mu^E(d) 
&\coloneqq& \dfrac{1}{N} \dsum_{i=1}^N \mbE_{\mX}[Y_i \mid f_i(\bX)=d].
\ee
Note that $\widetilde\mu^E(d)$ reduces to $\mu^E(d)$ under the correct specification of exposure mappings.
\citet{leung2022causal} assumed that the misspecified exposure mappings $f_i(\bX)$ depend on $\bX$ only via the $K$-neighborhoods of units ($K \geq 1$) in the interference graph $\bZ$ and established conditions under which the Horvitz-Thompson estimator is a consistent estimator of $\widetilde\tau^E(d_1, d_2)$.
\citet{savje2024causal} considered more general forms of misspecified exposures and established conditions in terms of the specification errors $Y_i - \mbE_{\mX}[Y_i\mid f_i(\bX) = d]$ to obtain variance bounds on Horvitz-Thompson estimators and establish consistency.

\section{Assumptions and Scope}
\label{sec:finite.super.population.inference}

The preceding sections demonstrate that there are many approaches to causal inference under interference.
It is useful to distinguish these approaches according to the assumptions made to achieve desirable properties of statistical procedures (e.g., consistency and asymptotic normality of estimators),
and the resulting scope of causal conclusions.
{\blues
We therefore review
\bi
\item assumptions that affect the scope of causal conclusions (Section \ref{sec:comparing.design.model.based.estimators});
\item the scope of causal conclusions based on fixed and random potential outcomes (Sections \ref{subsec:fixedgraph.fixedoutcome}--\ref{subsec:random});
\item the scope of causal conclusions based on a fixed or random interference graph (Section \ref{sec:fixed.versus.random.interference.graph}).
\ei}
Throughout Sections \ref{sec:comparing.design.model.based.estimators}--\ref{sec:fixed.versus.random.interference.graph},
we assume that observations of treatment assignments, 
outcomes,
and connections of all units in the population of interest are available.
The open problem of external validity in causal inference under interference---i.e.,
extending conclusions from a sample to a population (generalizability) or from one population to other populations (transportability)---is discussed in Section \ref{sec:external.validity}.

\subsection{Assumptions} 
\label{sec:comparing.design.model.based.estimators}

Beyond the assumptions required for identification of causal effects (Section \ref{sec:identification}),
additional assumptions are needed to ensure that statistical procedures possess desirable properties,
such as consistency and asymptotic normality of estimators.
At a high level, 
these assumptions seek to control interference.
We review some of the most widely used assumptions restricting interference in the fixed and random potential outcome frameworks.

\subsubsection{Assumptions: Interference Graph}
\label{assumption2}

The fixed and random potential outcome frameworks often restrict interference by restricting the sizes of neighborhoods.
A common assumption is that the number of neighbors $\sum_{j=1}^N z_{i,j}$ or the expected number of neighbors $\mbE\, \sum_{j=1}^N Z_{i,j}$ grows slower than $N-1$,
the maximum number of neighbors.
The specific requirements depend on many factors, 
including the causal effect of interest and the method of estimation.

\subsubsection{Assumptions: Potential Outcomes}
\label{assumption1}

A widely used assumption about potential outcomes is neighborhood exposure:
the potential outcomes of any unit $i$ are only affected by its own treatment assignment $x_i$ and the treatment assignments $x_j$ of its neighbors $j$ in the interference graph \citep[e.g.,][]{ugander2013graph,SuAi17, jagadeesan2020designs,leung2020treatment,forastiere2021identification,LiWa22,ogburn2024causal}.

While common, 
neighborhood exposure assumptions can be violated by contagion \citep{eckles2017design},
because contagion allows the treatment assignment $x_j$ of any unit $j$ to affect the potential outcomes of any other unit $i$ directly or indirectly,
regardless of whether $i$ and $j$ are neighbors in the interference graph. 
The violation of neighborhood exposure assumptions by contagion is demonstrated in Figures \ref{fig:con.pop} and \ref{graph.exposure}.
Violations of neighborhood exposure assumptions may not affect estimators of direct causal effects,
but can bias estimators of indirect causal effects \citep{BhSc25}.

There have been recent attempts to relax neighborhood exposure assumptions.

First,
in the literature based on fixed potential outcomes $\by(\cdot)$,
\citet{leung2022causal} introduces an approximate neighborhood interference (ANI) assumption under which the treatment assignment $x_j$ of any unit $j$ can affect the potential outcomes $y_i(\bx)$ of any other unit $i$, 
but the strength of interference decays as the geodesic distance between $i$ and $j$ increases;
recall that the geodesic distance between $i$ and $j$ is the length of the shortest path connecting them.
\citet{yuan2021motif} study exposure mappings that can be functions of subgraphs (e.g., stars, triangles) rather than neighbors.

Second,
in the literature based on random potential outcomes $\bY(\cdot)$, 
\citet{belloni2025neighborhood} consider the case when the units $i$ in the population can experience interference differently via $m_i$-hop neighborhoods,
but assume independence among potential outcomes,
ruling out contagion (Sections \ref{sec:causation.versus.correlation} and \ref{sec:capturing.contagion}).
\citet{TTFuSh21} point out that Markov random fields \citep{rue2005gaussian,La96} allow the potential outcomes $Y_i(\bx)$ of units $i$ to depend on the treatment assignments $x_j$ of all other units $j$ who are connected to $i$ by a finite path.

\begin{figure}
\begin{center}
      \tikz{ %
        \node[\obs] (y1) {$1$} ; %
        \node[\obs, right=of y1] (y2) {$2$} ; %
        \node[\obs, right=of y2] (y3) {$\ldots$} ; %
        \node[\obs, right=of y3] (y4) {$N$} ; %
        \edge[-, line width=1pt]{y1}{y2} ; %
        \edge[-, line width=1pt]{y2}{y3} ; %
        \edge[-, line width=1pt]{y3}{y4} ; %
}                                                                                                       
\end{center}          
\caption{\label{fig:con.pop}
A simple example of an interference graph $\bZ$ with $N$ units,
where $Z_{i,j} \coloneqq 1$ if $|i - j| = 1$ and $Z_{i,j} \coloneqq 0$ otherwise.
}
\label{fig:placeholder}
\end{figure}
\begin{figure}
\begin{center}
      \tikz{ %
        \node[\obs] (y1) {$Y_1$} ; %
        \node[\obs, right=of y1] (y2) {$Y_2$} ; %
        \node[\obs, below=of y1, yshift=0cm] (h1) {$X_1$} ; %
        \node[\obs, below=of y2, yshift=0cm] (h2) {$X_2$} ; %
        \node[\obs, right=of h2, yshift=0cm] (h3) {$\cdots$} ; %
        \node[\obs, right=of h3, yshift=0cm] (h4) {$X_N$} ; %
        \node[\obs, right=of y2, yshift=0cm] (y3) {$\cdots$} ; %
        \node[\obs, right=of y3, yshift=0cm] (y4) {$Y_N$} ; %
        \edge[color=MidnightBlue, line width=1pt]{h2} {y3} ; %
        \edge[color=black, line width=1pt]{h4} {y4} ; %
        \edge[color=MidnightBlue, line width=1.15pt]{h3} {y4} ; %
        \edge[color=MidnightBlue, line width=1.15pt]{h4} {y3} ; %
        \edge[color=black, line width=1pt]{h1} {y1} ; %
        \edge[color=black, line width=1pt]{h3} {y3} ; %
        \edge[color=MidnightBlue, line width=1.15pt]{h1} {y2} ; %
        \edge[color=MidnightBlue, line width=1.15pt]{h3} {y2} ; %
        \edge[color=black, line width=1pt]{h2} {y2} ; %
        \edge[color=MidnightBlue, line width=1pt]{h2} {y1} ; %
        \edge[color=black, line width=1pt]{h2} {y2} ; %
        \edge[-, color=orange, line width=1.25pt]{y1}{y2} ; %
        \edge[-, color=orange, line width=1.25pt]{y2}{y3} ; 
        \edge[-, color=orange, line width=1.25pt]{y3}{y4} ; 
      }
\caption{\label{graph.exposure}
Violation of neighborhood exposure assumptions by contagion,
assuming that the interference graph is given by Figure \ref{fig:con.pop}.
Contagion allows any $X_i$ to affect any $Y_j$ either directly or indirectly,
regardless of whether $i$ and $j$ are neighbors in the interference graph.
}
\end{center}
\end{figure}

\subsection{Scope: Fixed Potential Outcomes}
\label{subsec:fixedgraph.fixedoutcome}

Approaches based on fixed potential outcomes have at least two advantages.
First,
if potential outcomes and interference graphs $(\by(\cdot), \bz)$ are fixed,
investigators do not need to make assumptions about the process generating $(\by(\cdot), \bz)$.
Second,
differences in potential outcomes 
\beno 
y_i(x_i = 1; \bx_{-i}) - y_i(x_i = 0; \bx_{-i})\s
\\
y_j(x_i = 1; \bx_{-i}) - y_j(x_i = 0; \bx_{-i})
\ee
can be attributed to differences in the treatment assignment $x_i$ of unit $i$,
because the sole source of randomness stems from the treatment assignments $\bX$.

While the fixed potential outcome framework does not make assumptions about the process generating $(\by(\cdot), \bz)$, 
statistical procedures may not possess desirable properties in the presence of interference unless restrictions are imposed on the set of possible $(\by(\cdot), \bz)$:
e.g.,
consistency of estimators may require restricting $(\by(\cdot), \bz)$ to a subset of potential outcomes $\mY_s(\cdot)$ satisfying neighborhood exposure and a subset of interference graphs $\mZ_s$ with bounded numbers of neighbors,
although these assumptions can be relaxed.
Having said that,
some properties of estimators based on the fixed potential outcome framework carry over to the random potential outcome framework.
We discuss them in Section \ref{fixed-to-random}.

\subsection{From Fixed to Random Potential Outcomes}
\label{fixed-to-random}

Since causal effects based on fixed potential outcomes and interference graphs $(\by(\cdot),\bz)$ can be viewed as conditional causal effects given $(\bY(\cdot),\bZ) = (\by(\cdot),\bz)$,
some properties of statistical procedures extend from fixed to random potential outcomes,
but not all of them do.

We first show that unbiasedness extends from the fixed potential outcome framework to the random potential framework,
although some restrictions on the set of $(\by(\cdot), \bz)$ may be required.
An example is the Horvitz-Thompson estimator $\widehat\tau_{HT}^D$ discussed in Section \ref{sec:design.based.estimators}:
the unbiasedness of $\widehat\tau_{HT}^D$ as an estimator of the direct causal effect $\tau_{\mX}^D$ based on fixed $(\by(\cdot), \bz)$ extends to the direct causal effect $\tau_{\mX,\mY(\cdot),\mZ}^D$ based on random $(\bY(\cdot), \bZ)$.
Estimators of indirect and total causal effects,
such as $\tau_{\mX}^I$ and $\tau_{\mX} = \tau_{\mX}^D + \tau_{\mX}^I$,
possess the same property,
but may require restrictions on the set of $(\by(\cdot), \bz)$.
Consider any causal effect $\theta_\mX$ based on fixed $(\by(\cdot), \bz)$.
Since $\theta_\mX$ can depend on $(\by(\cdot), \bz)$,
we henceforth write $\theta_\mX(\by(\cdot),\bz)$ instead of $\theta_\mX$.
Let $\widehat\theta$ be any unbiased estimator of $\theta_\mX(\by(\cdot),\bz)$:
\beno
\mbE_{\mX\mid\mY(\cdot),\mZ}\, [\,\widehat\theta\mid (\bY(\cdot),\bZ)=(\by(\cdot),\bz)] 
&\;=\;& \theta_\mX(\by(\cdot),\bz).
\ee
Note that $\widehat\theta$ may not be an unbiased estimator of
$\theta_\mX(\by(\cdot),\bz)$ for all $(\by(\cdot), \bz) \in \mY(\cdot)\times \mZ$:
e.g.,
\citet{LiWa22} present an unbiased Horvitz-Thompson estimator for the total causal effect $\tau_\mX$ based on the assumption that the potential outcomes $\by(\cdot)$ are contained in a subset of potential outcomes $\mY_s(\cdot)$ satisfying neighborhood exposure.
In other words,
the unbiasedness of $\widehat\theta$ as an estimator of $\theta_\mX(\by(\cdot),\bz)$ may be restricted to elements $(\by(\cdot), \bz)$ of a proper subset $\mY_s(\cdot) \times \mZ_s$ of $\mY(\cdot) \times \mZ$.
If the support of $(\bY(\cdot),\bZ)$ is contained in $\mY_s(\cdot)\times \mZ_s$,
the unbiasedness of $\widehat\theta$ as an estimator of $\theta_\mX(\by(\cdot),\bz)$ for all $(\by(\cdot), \bz) \in \mY_s(\cdot) \times \mZ_s$ implies that 
\beno
\label{unbiasedness}
\mbE_{\mX,\mY(\cdot),\mZ}\; \widehat\theta
&\,=\,& \mbE_{\mY(\cdot),\mZ}\, [\, \mbE_{\mX\mid\mY(\cdot),\mZ}\, [\,\widehat\theta\mid \bY(\cdot),\bZ]]\s 
\\
&\,=\,& \mbE_{\mY(\cdot),\mZ}\; \theta_\mX(\bY(\cdot),\bZ)\s 
\\
&\,\eqqcolon\,& \theta_{\mX,\mY(\cdot),\mZ}.
\ee
Thus,
if $\widehat\theta$ is an unbiased estimator of the causal effect $\theta_\mX(\by(\cdot),\bz)$ based on fixed $(\by(\cdot),\bz) \subseteq \mY_s(\cdot)\times \mZ_s$,
then it is an unbiased estimator of the causal effect $\theta_{\mX,\mY(\cdot),\mZ}$ based on random $(\bY(\cdot),\bZ)$,
provided that the support of $(\bY(\cdot),\bZ)$ is contained in $\mY_s(\cdot)\times \mZ_s$.

While unbiasedness extends from fixed to random potential outcomes subject to restrictions on the set of $(\by(\cdot), \bz)$,
the coverage probabilities of confidence intervals may not.
To demonstrate,
consider any $\alpha \in (0, 1)$ and let $\mathscr{S}$ be an interval estimator of $\theta_\mX(\by(\cdot),\bz)$,
provided that $(\by(\cdot),\bz)\in\mY_s(\cdot)\times \mZ_s$.
If
\be
\label{condition00}
\mbP_{\mX\mid\mY(\cdot),\mZ}(\theta_\mX(\bY(\cdot),\bZ) \in \mathscr{S} | (\bY(\cdot),\bZ)=(\by(\cdot),\bz)) > 1-\alpha\s\\
\qquad\qquad\qquad\text{for all $(\by(\cdot),\bz)\in \mY_s(\cdot)\times \mZ_s$,}
\ee
then
\beno
\mbP_{\mX,\mY(\cdot),\mZ}(\theta_\mX(\bY(\cdot),\bZ) \in \mathscr{S})
> 1-\alpha,
\ee
provided that $(\bY(\cdot),\bZ)$ is generated from a distribution with support contained in $\mY_s(\cdot)\times \mZ_s$.
That said,
even when the condition in Equation \eqref{condition00} is satisfied, 
$\mbP_{\mX,\mY(\cdot),\mZ}(\theta_{\mX,\mY(.),\mZ} \in\mathscr{S})$ may be less than $1-\alpha$,
because confidence intervals for $\theta_{\mX,\mY(\cdot),\mZ}$ will depend on the variability of $\theta_{\mX}(\bY(\cdot),\bZ)$ around its mean $\theta_{\mX,\mY(\cdot),\mZ}$.

\subsection{Scope: Random Potential Outcomes}
\label{subsec:random}

While the fixed potential outcome framework is sufficient in many applications,
the random potential outcome framework helps expand the scope of causal conclusions when the data fall short \citep[e.g.,][]{ToKa13,leung2020treatment,TTFuSh21,LiWa22,FoMeWuAi22,ClHa24,BhSe24}.
We give two examples in which the data fall short and the random potential outcome framework has added value.

The first example concerns heterogeneous treatment effects \citep[e.g.,][]{bong2025heterogeneous,yuan2021motif,adhikari2025inferring,bargagli2025heterogeneous}.
If no assumptions are made about the process generating the potential outcomes $y_i(\cdot)$ of units $i$,
it may not be possible to estimate individual treatment effects,
because one potential outcome of unit $i$ is observed while its other $2^{N-1}$ potential outcomes are missing.
By contrast,
if investigators are willing to make assumptions about the process generating potential outcomes $y_i(\cdot)$,
individual treatment effects can be estimated by pooling strength across similar units \citep[e.g., units with the same number of neighbors:][]{bong2025heterogeneous}.

The second example concerns external validity,
including generalizability and transportability.
The problems of generalizability and transportability are fundamental problems in all areas of statistics.
A classic example is survey sampling \citep{hartley1975super,Th12}: 
if a sample of size $n < N$ is collected from a finite population of size $N$ with outcomes $y_1, \ldots, y_N$,
the sample mean can be used to infer the population mean,
\bi 
\item[(a)] without making assumptions about the process generating $y_1, \ldots, y_N$ (finite population inference),
\vspace{.1cm}
\item[(b)] or by assuming that $y_1, \ldots, y_N$ are realizations of random variables $Y_1, \ldots, Y_N$ governed by a probability law $\mbP$ (super population inference).
\ei 
In finite population inference,
all conclusions are limited to $y_1, \ldots, y_N$.
If $y_1, \ldots, y_N$ represent the political preferences of eligible voters in the U.S.,
finite population inference is sufficient:
inferences about the political preferences of U.S.\ voters in the future or populations in other countries may neither be desirable nor possible.
Super population inference can be useful if investigators wish to transport inferences from one population of interest (e.g., one company, school, or country) to other populations of interest (e.g., other companies, schools, or countries).
To extend conclusions from the study population to other populations,
there must be a relationship between the study population and the other populations.
A possible relationship between these populations is that potential outcomes in these populations are generated by probability laws that are similar in a well-defined sense.
While complicated by spillover,
contagion,
and other complex real-world processes in connected populations,
the same idea applies to causal inference under interference:
super population approaches based on random potential outcomes governed by a probability law help expand the scope of conclusions to other populations with similar characteristics.
In the presence of interference, 
external validity is an open problem.
We discuss external validity under interference in Section \ref{sec:external.validity}.

Having said that,
super population approaches based on random potential outcomes \citep[e.g.,][]{ToKa13,leung2020treatment,TTFuSh21,LiWa22,FoMeWuAi22,ClHa24,BhSe24} require more assumptions about the process generating potential outcomes and are sensitive to violations of those assumptions,
though nonparametric approaches help alleviate such concerns \citep[e.g.,][]{ogburn2024causal,bong2025heterogeneous}. 

\subsection{Scope: Fixed versus Random Interference Graph}
\label{sec:fixed.versus.random.interference.graph}

There are two legitimate approaches to dealing with an underlying interference graph:
\bi
\item Some approaches view the interference graph as fixed,
while others view it as random but condition on it.
Such approaches do not require assumptions about the graph-generating process and are not prone to misspecification of it,
but may require stronger conditions on the interference graph in the form of sparsity.
For example, 
\citet{LiWa22} argue that the Horvitz-Thompson estimator $\widehat\tau_{HT}^D$ of the direct causal effect $\tau^D_{\mX}$ achieves the rate of convergence $1/\sqrt{N}$ provided that the interference graph is fixed and sparse,
in the sense that the number of neighbors of each unit is bounded.
\vspace{.05cm}
\item By contrast,
when the interference graph is random,
conclusions can be extended from sparse to dense interference graphs:
e.g.,
\citet{LiWa22} show that the Horvitz-Thompson estimator $\widehat\tau_{HT}^D$ of the direct causal effect $\tau^D_{\mX}$ achieves the rate of convergence $1/\sqrt{N}$ without bounded numbers of neighbors,
provided that the interference graph is generated by a graphon model with unbounded degrees and the potential outcomes are smooth functions of treatment assignments.
That said,
conclusions based on random interference graphs are sensitive to misspecification of the graph-generating process.
\ei

An additional consideration is that average potential outcomes and causal effects may depend on the structure of the interference graph in the presence of interference,
and conclusions may not be extendable across interference graphs when the set of possible interference graphs exhibits high variability.
In the running example,
the presence of influencers with millions of followers can affect average potential outcomes of interventions:
in the absence of influencers,
average potential outcomes can be smaller than in the presence of influencers.
In other words,
if some of the possible interference graphs include influencers while others do not,
fixing or conditioning on a known interference graph without influencers may not allow investigators to extend conclusions to other possible interference graphs with influencers.

\vspace{.1cm}

\begin{figure}
\begin{center}
\includegraphics[width=0.7\linewidth]{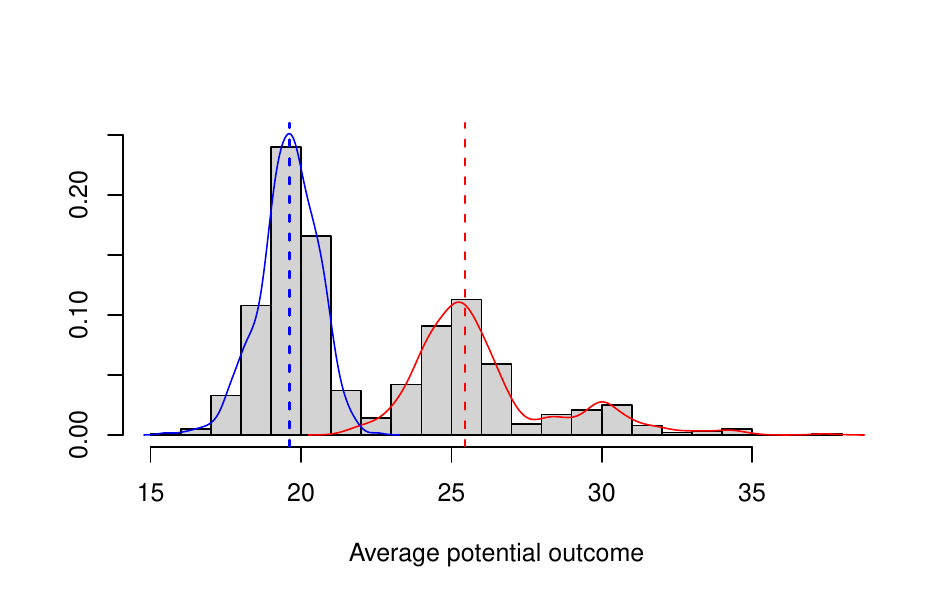}
\end{center}
\caption{
Average potential outcome $\mu(\bx^\star)$ based on 1,000 simulated data sets with $N = $ 1,000 units.
The blue-colored component of the distribution represents the average outcome based on interference graphs without treated superstars,
whereas the red-colored component represents the average outcome based on interference graphs with at least one treated superstar.
The dashed vertical lines represent the medians of these components.}
\label{fig:two_world}
\end{figure}

{\em Simulations.}
To demonstrate that high variability can make it challenging to extend conclusions across interference graphs,
we generate 1,000 data sets with $N = $ 1,000 units.
In each data set,
the first 500 units are assigned treatment and the other 500 units are not.
The resulting treatment assignments are fixed and denoted by $\bx^\star$.
The interference graph $\bZ$ is generated by a $\beta$-model \citep{ChDiSl11},
one of the simplest models for generating networks consisting of many units with few neighbors and some units with many neighbors (e.g., influencers with many followers).
The $\beta$-model generates $\bZ$ by sampling
\beno 
Z_{i,j} 
&\ind& \text{Bernoulli}(P_{i,j}),\;
& P_{i,j}
&\coloneqq& \dfrac{\exp(\omega_i + \omega_j)}{1+\exp(\omega_i + \omega_j)}.
\ee
The weights $\omega_i \in \mbR$ and $\omega_j \in \mbR$ can be interpreted as the propensities of units $i$ and $j$ to connect with others.
We generate these propensities by first sampling
\beno
U_i 
&\iid& \text{Bernoulli}(1/1000),
\ee
and then setting 
\beno
\omega_i 
&\coloneqq& 
\begin{cases}
-2 & \mbox{if } U_i = 0\\
20 & \mbox{if } U_i = 1.
\end{cases}
\ee
A unit $i$ with $U_i = 1$ is interpreted as a superstar,
in the sense that $i$ has a high propensity $\omega_i = 20$ to connect with others and will therefore be the center of a large star in the interference graph \citep{La96}.
The $\beta$-model creates two worlds:
a world without superstars and a world with $1, 2, \ldots$ superstars.
The potential outcomes $\by(\bx^\star)$ are realizations of $\bY(\bx^\star) \mid (\bU, \bZ) = (\bu, \bz)$ generated from a multivariate Gaussian with conditional expectations of\, $Y_i(\bx^\star) \mid (U_i, \bY_{-i}(\bx^\star), \bZ) = (u_i, \by_{-i}(\bx^\star), \bz)$ satisfying
\beno
\label{natural2}
\mbE_{\mY_i(\bx^\star) \mid \mathscr{U}_i, \mY_{-i}(\bx^\star), \mZ}[Y_i(\bx^\star) \mid (u_i, \by_{-i}(\bx^\star), \bz)]\s
\\
\coloneqq 
\beta\, (1 + \epsilon\, u_i)\, x_i^\star
\hspace{-0.05cm}
+ 
\hspace{-0.05cm}
\gamma\, \dfrac{\sum_{j=1}^N x_j^\star\, z_{i,j}}{\sum_{j=1}^N z_{i,j}}
\hspace{-0.05cm}
+ 
\hspace{-0.05cm}
\delta\, \dfrac{\sum_{j=1}^N y_j(\bx^\star)\, z_{i,j}}{\sum_{j=1}^N z_{i,j}},
\ee
with $\beta = 2$,\,
$\gamma=2$,\,
$\delta = 0.9$,\,
$\epsilon=10$, 
and conditional variance $N/\,\sum_{j=1}^N\,z_{i,j}$.
The idea of the potential outcome model is that the expected potential outcomes of treated superstars $j$ ($u_j = 1$ and $x_j^\star = 1$) are boosted by $\epsilon = 10$,
whereas the expected potential outcomes of units $i$ connected to superstars $j$ are boosted by $\delta\, y_j(\bx^\star) / \sum_{j=1}^N z_{i,j}$:
e.g.,
in the running example,
if advertisers target influencers with millions of followers with advertisements of designer clothes and offer them financial incentives for purchasing designer clothes,
then other teenagers may likewise purchase designer clothes.
Figure \ref{fig:two_world} shows the average potential outcome
\beno
\mu(\bx^\star)
&\coloneqq& \dfrac1N \dsum_{i=1}^N Y_i(\bx^\star)
\ee
based on the 1,000 simulated data sets.
The blue-colored distribution on the left-hand side of Figure \ref{fig:two_world} shows the average potential outcomes in a world without treated superstars,
whereas the red-colored distribution on the right-hand side shows the average potential outcomes in a world with one or more treated superstars.

These simulation results highlight two salient features of causal inference under interference:
\bi
\item[1.] Causal effects based on linear combinations of potential outcomes can depend on the structure of the interference graph (e.g., the absence or presence of superstars  or communities).
\vspace{.05cm}
\item[2.] If another interference graph were observed,
causal conclusions could be different:
e.g.,
if the interference graph changed over time---as many real-world networks do---conclusions could be invalidated.
\ei
These findings suggest the need to understand what would happen if another interference graph were observed, 
and how different the resulting conclusions might be.
Well-chosen models of interference graphs can help,
but there is a risk of misspecifying the graph-generating process.

\section{Directions for Future Research}
\label{sec:discussion}

We conclude with a selection of possible directions for future research.

\hide{

\subsection{Detecting Contagion}
\label{sec:detecting.contagion}

The simulation results in Section \ref{sec:simulation} demonstrate that existing tests can detect interference arising from spillover.
The same simulation results reveal that tests based on neighborhood exposure assumptions may be unable to detect interference arising from contagion,
because contagion can violate neighborhood exposure assumptions.
{\blues
A promising avenue for future research is the development of tests that do not rely on neighborhood exposure assumptions,
which can be accomplished in the fixed or random potential outcome framework.}

}

\subsection{The Role of Networks}
\label{sec:network.effect}

Throughout the survey,
we have used interference graphs to represent the network of channels that allow the treatment assignments of units to affect the potential outcomes of other units.
Those channels may reflect underlying economic, 
social,
or physical relationships.
Treatment assignments can affect those relationships,
and those relationships in turn enable the effects of treatments to propagate throughout connected populations.

First,
in the bulk of the literature on interference, the network has been treated as a pre-treatment variable that can affect treatment assignments and outcomes.
However, 
the network itself might be affected after treatments are assigned, which in turn might affect outcomes:
e.g.,
\citet{gao2024endogenous} studies causal effects when treatment assignments affect networks along with outcomes.

Second,
networks are complex and high-dimensional objects with a large number of features (e.g., degrees, cycles, stars, cliques, communities, and structural holes), and
many features can affect causal effects.
To date,
networks have often been reduced to low-dimensional statistics.
For example,
many exposure mappings reduce the network and its effect on potential outcomes to 
\bi 
\item the number of neighbors $\sum_{j=1}^N z_{i,j}$,
\item the number of treated neighbors $\sum_{j=1}^N x_j\, z_{i,j}$.
\ei 
Such assumptions imply that each neighbor $j$ affects the expected potential outcome of unit $i$ by the same amount,
regardless of the position of $j$ in the network (e.g., whether $j$ is a superstar, 
a member of a desirable community, 
or a ``bridge'' between communities).
\citet{yuan2021motif} use decision trees to identify relevant network features.
That said,
the understanding of how causal effects are affected by the structure of networks is incomplete,
and more insight into the impact of network structure on causal effects could help improve interventions in connected populations and predictions.

\subsection{Time-Indexed Data}
\label{subsec:temporal}

Causal processes are temporal processes:
the effects of treatment,
spillover,
and contagion cannot materialize at the same time.
{\blues
In the running example,
the effect of treatment (advertisements) on outcomes (purchases of designer clothes) and the effects of interference (friends of exposed teenagers purchasing designer clothes) can take weeks or months to materialize.}

If there are no repeated observations of treatment assignments $\bX$ and outcomes $\bY$,
then $\bX$ and $\bY$ can be viewed as a single realization from the stationary distribution of a time-indexed stochastic process \citep{LaRi02,OgShLe20}.
Otherwise,
when there are repeated observations of outcomes,
it is possible to consider a time-indexed stochastic process as a model of the causal process of interest.

\citet{eckles2017design} and \citet{La14} have taken first steps toward causal inference under interference based on repeated observations of outcomes.
For example,
\citet{eckles2017design} study outcomes observed at time points $0, 1, \ldots, T$ ($T \geq 1$) generated by a time-indexed stochastic process satisfying
\beno
Y_{i,t} 
&=& f_{i,t}(X_i,\, Y_{i,t-1},\, \{Y_{j,t-1}:\, Z_{i,j}=1\},\, \bm{U}_t),
\ee
where $f_{i,t}$ is a function of the treatment assignment $X_i$ of unit $i$,
the past outcome $Y_{i,t-1}$ of unit $i$,
the past outcomes $Y_{j,t-1}$ of other units $j$ connected to unit $i$,
and 
exogenous errors $\bm{U}_t \in \mR^N$.
\citet{eckles2017design} consider estimating the causal effect
\beno
\tau^T(\bm{1}_N, \bm{0}_N)
&\coloneqq& \dfrac{1}{N} \dsum_{i=1}^N \mbE_{\mY(\cdot)\mid \mZ}\, [Y_{i,T}(\bm{1}_N) - Y_{i,T}(\bm{0}_N)],
\ee
which resembles the causal effect $\tau(\bm{1}_N, \bm{0}_N)$ introduced in Section \ref{sec:causal.effects.fixed},
but compares potential outcomes $\bY_T(\bm{1}_N)$ and $\bY_T(\bm{0}_N)$ at time $T$. 
To estimate $\tau^T(\bm{1}_N, \bm{0}_N)$ under exposure mappings, 
one can estimate a form of the difference-in-exposures effect $\tau^E_{\mY(\cdot)\mid \mZ}(d_1,d_2)$,
where the exposures $d_1$ and $d_2$ correspond to treatment assignments $\bm{1}_N$ and $\bm{0}_N$, 
respectively \citep[see][]{ugander2013graph}.
In the absence of exposure mappings,
\citet{eckles2017design} show that $\tau^T(\bm{1}_N, \bm{0}_N)$ is close to a variant of the direct effect $\tau^D_{\mX,\mY(\cdot)\mid\mZ}$ under graph cluster randomization,
and can hence be estimated by any (consistent) estimator of $\tau^D_{\mX,\mY(\cdot)\mid\mZ}$.
\citet{La14} considered a more general data-generating process,
with both treatment assignments $X_{i,t}$ and outcomes $Y_{i,t}$ evolving over time:
\beno
X_{i,t} 
&=& f^{\mX}_t(s^{\mX}_{i,t}(\pa(X_{i,t})),\, U^{\mX}_{i,t})\s
\\
Y_{i,t}  
&=& f_t^{\mY}(s^{\mY}_{i,t}(\pa(Y_{i,t})),\, U^{\mY}_{i,t}),
\ee
where $\pa(X_{i,t})$ and $\pa(Y_{i,t})$ consist of all past observations that can affect $X_{i,t}$ and $Y_{i,t}$,
which can be viewed as the parents of $X_{i,t}$ and $Y_{i,t}$ in a directed acylic graph \citep{La96}.
The functions $s^{\mX}_{i,t}(\cdot)$ and $s^{\mY}_{i,t}(\cdot)$ project past observations into a low-dimensional space,
facilitating inference.
\citet{La14} develops targeted maximum likelihood estimators to estimate the average expected potential outcomes at the final time point $T$.
More recent work by \citet{shirani2024message} deals with unknown interference in temporal settings using message-passing.

Despite these advances,
causal inference under interference based on repeated observations is an open area of research with great promise.

\subsection{Space- and Time-Indexed Data}
\label{sec:spatial.interference}

A related area of research with many open questions is spatial interference \cite{reich2021review}:
e.g.,
\citet{ChBaKuMaPe22} study the effect of conflict on forest loss in Colombia \citep{ChBaKuMaPe22},
while \citet{papadogeorgou2022causal} study the effect of airstrikes on the insurgency in Iraq.
In both cases,
there is spatial interference,
in the sense that the outcomes of a region can be affected by the treatments or outcomes of neighboring regions due to spillover or contagion.
Spatial interference can be viewed as a special case of network interference,
where the interference graph has weighted connections and the weights of connections may be functions of the spatial distance $d_{i,j}$ between units $i$ and $j$,
e.g.,
$Z_{i,j} \coloneqq \exp(-d_{i,j})$.
We refer the interested reader to \citet{wang2025design} and \citet{leung2022rate} for statistical theory,
and to \citet{ChBaKuMaPe22} and \citet{zigler2020bipartite} and \citet{papadogeorgou2022causal} for case studies 
with spatial interference.
Another related branch of literature in economics is concerned with interventions in market settings with spatio-temporal structure,
such as housing markets.
We refer interested readers to \citet{tu2026panel} and references therein.

\subsection{More than Means}

The bulk of literature concerned with random potential outcomes focuses on the effect of treatments on the means of potential outcomes.
In principle,
treatments can affect other features of the distribution of random potential outcomes:
e.g.,
treatments may affect the spread of the distribution in addition to its mean,
as well as its tails.
A wholesome assessment of the causal effect of treatments on the distribution of potential outcomes constitutes an interesting direction for future research.

\hide{

\subsection{Potential Outcomes: Objects}
\label{sec:random.objects}

To the best of our knowledge,
potential outcomes that are objects in a metric space constitutes an unchartered area of research in causal inference under interference.
Examples include functional data and manifolds,
such as the development curves of children that may be affected by other children connected by kinship or friendship.

\vspace{.1cm}

}

\subsection{External Validity}
\label{sec:external.validity}

An open problem in causal inference under interference is the external validity of causal conclusions.
A well-designed experiment ensures internal validity,
in the sense that the causal conclusions are valid in the sample on which the conclusions are based.
The problem of external validity concerns the question of whether---and how---the causal conclusions can be generalized from the sample to the population from which the sample is taken (generalizability) \cite{lesko2017generalizing} or other populations of interest (transportability) \citep{PeBa14}.
In the running example,
advertisers may be interested in estimating the average increase in purchases due to advertisements in the worldwide population of teenagers with purchasing power based on a sample.

First steps toward external validity under interference are taken by \citet{leung2020treatment} and \citet{ReLuChMC24},
but external validity remains an open problem.
For example,
\citet{leung2020treatment} requires the sample size $n$ to be of the same order as the population size $N$.
An additional issue is dependence among potential outcomes due to contagion and other sources,
which implies that potential outcomes of sampled units can be affected by treatment assignments and potential outcomes of unsampled units.
How to generalize findings in such scenarios without sacrificing scalability and theoretical guarantees is an unsolved problem.

\subsection{Optimal Interventions}
\label{sec:optimal.interventions}
A promising area of research is optimal interventions in connected populations.
If investigators have established that interventions affect outcomes of interest,
policy makers may wish to maximize the effect of interventions by finding an optimal treatment assignment rule based on individual characteristics of units \citep{zhang2025individual,viviano2024policy,XuLuSo24,GlLaVo25}.

Assuming that potential outcomes $\bY(\bx)$ are random,
the average expected potential outcome resulting from treatment assignments $\bx$ can be defined as
\beno
\label{social.welfare}
\mu_{\mY(\cdot)\mid\mC,\mZ}(\bx) 
&\coloneqq& \dfrac{1}{N}\dsum_{i=1}^N \mbE_{\mY(\cdot)\mid\mC,\mZ}[Y_i(\bx) \mid (\bm{C}, \bZ) = (\bm{c}, \bz)],
\ee
where $\mbE_{\mY(\cdot)\mid\mathscr{C},\mZ}[Y_i(\bx) \mid (\bm{C}, \bZ) = (\bm{c}, \bz)]$ is the expected potential outcome $Y_i(\bx)$ of unit $i$ when the treatments assignments are $\bx \in \mX$ conditional on covariates $\bm{c} \in \mC\coloneqq\mC_0^N \subseteq \mR^{N \times p}$ ($p \geq 1$) and interference graph $\bz \in \mZ$.
Policy makers may then be interested in finding an optimal treatment assignment rule $\pi^\star$ in a set of possible assignment rules $\Pi\subset\{\pi:\pi \text{ is a map from }\mC_0\text{ to }\{0,1\}\}$:
\beno
\pi^\star
&\in& \argmax\limits_{\pi \in \Pi}\, \mu_{\mY(\cdot)\mid\mC,\mZ}(\pi(c_1),\ldots,\pi(c_N)).
\ee
In practice,
restrictions are imposed on $\Pi$ and 
$\mu_{\mY(\cdot)\mid\mC,\mZ}(\bx)$ can be replaced by an estimator $\widehat\mu_{\mY(\cdot)\mid\mC,\mZ}(\bx)$,
such as a doubly robust estimator \citep{viviano2024policy}.
Two examples are maximizing welfare in economics \citep{viviano2024policy} or rewards in reinforcement learning in connected populations \citep{GlLaVo25}.

\section*{Funding}
We acknowledge support in the form of U.S.\ National Science Foundation award NSF DMS-2515763 and U.S.\ Army Research Office award ARO W911NF-21-1-0335.

\section*{Acknowledgements}

We are indebted to three anonymous referees whose numerous constructive suggestions greatly improved the manuscript.
In addition,
we would like to thank Samuel Baugh, 
Abhishek Chakraborty,
Ju Hyun Oh,
Hyebin Song,
and the causal inference group at The Pennsylvania State University for helpful discussions.

\bibliographystyle{chicagoa} 
\bibliography{base, causal}

\end{document}